\pdfoutput=1

\def\jpsi{J\kern-0.15em/\kern-0.15em\psi\kern0.15em}

\newcommand{\lsim}{\lesssim}
\newcommand{\gsim}{\gtrsim}
\def\nl{\hfill\break}
\documentclass[12pt,a4paper]{article}

\usepackage{ifthen} 

\usepackage{graphicx}  
\usepackage{color}
\newlength\typewidth
\newlength\lfskip
\newboolean{articletitles}
\setboolean{articletitles}{true} 

\usepackage[numbers]{natbib}

\usepackage[breaklinks]{hyperref}    
\usepackage[all]{hypcap} 
\usepackage{mciteplus}

\usepackage{amssymb}

\begin{document}

\begin{titlepage}

\quad
\vspace*{3.0cm}

{\bf\boldmath\huge
\begin{center}
Multiquark States
\end{center}
}

\begin{center}
Marek Karliner,$^1$ Jonathan L. Rosner,$^2$ and Tomasz Skwarnicki$^3$
\bigskip\\
{\it\footnotesize
$ ^1$School of Physics and Astronomy, Tel Aviv University, Tel Aviv 69978, Israel; \\email: marek@post.tau.ac.il\\
$ ^2$Enrico Fermi Institute and Department of Physics, University of Chicago, 5620 S. Ellis Avenue, Chicago, IL 60637, USA; \\email: rosner@hep.uchicago.edu\\
$ ^3$Department of Physics, Syracuse University, Syracuse, NY 13244, USA; \\e-mail: tskwarni@syr.edu\\
}
\quad\\
\today
\end{center}

\vspace{\fill}

\begin{abstract}
\noindent
Why do we see certain types of strongly interacting elementary particles and
not others?  This question was posed over 50 years ago in the context of the
quark model.  M. Gell-Mann and G. Zweig proposed that
the known mesons were $q \bar q$ and baryons $qqq$, with quarks known at the
time $u$ (``up''), $d$ (``down''), and $s$ (``strange'') having charges
(2/3,--1/3,--1/3).  Mesons and baryons would then have integral charges.
Mesons such as $qq \bar q \bar q$ and baryons such as $qqqq \bar q$ would also
have integral charges. Why weren't they seen?  They {\it have} now been seen,
but only with additional heavy quarks and under conditions which tell us a lot
about the strong interactions and how they manifest themselves.
The present article describes recent progress in our understanding of such
``exotic'' mesons and baryons.

\end{abstract}

\vspace*{1.1cm}
\vfill

\begin{center}
{\it  To be submitted to Annual Review of Nuclear and Particle Science.}
\end{center}

\end{titlepage}

\newpage

\tableofcontents

\section{INTRODUCTION}
\label{sec:introduction}

Why do we see certain types of elementary particles and not others?  This
question was posed over 50 years ago in the context of the quark model
\cite{GellMann:1964nj,*Zweig:1981pd}.  M. Gell-Mann and G.~Zweig proposed that
the known mesons were $q \bar q$ and baryons $qqq$, with quarks known at the
time $u$ (``up''), $d$ (``down''), and $s$ (``strange'') having charges
(2/3,--1/3,--1/3).  Mesons and baryons would then have integral charges.
Mesons such as $qq \bar q \bar q$ and baryons such as $qqqq \bar q$ would also
have integral charges. Why weren't they seen?  They {\it have} now been seen,
but only with additional heavy quarks and under conditions which tell us a lot
about the strong interactions and how they manifest themselves.
The present article describes recent progress in our understanding of such
``exotic'' mesons and baryons.

After some introductory words on early multiquark states, the quark model, and
quantum chromodynamics (QCD), we discuss light multiquark candidates in
Sec.~\ref{sec:lighthadrons}, heavy-light multiquark candidates in
Sec.~\ref{sec:heavylighthadrons}, and heavy quarkonium-like multiquark
candidates in Sec.~\ref{sec:quarkoniumlike}.  We treat states beyond those
detected at present in Sec.~\ref{sec:beyonddetected}, and summarize in
Sec.~\ref{sec:summary}.

\subsection{Nucleons And Their Molecules}
\label{sec:nucleons}

The symmetries of the strong interactions have a long history, starting with
isotopic spin ({\it isospin}) which recognized the similarity of the neutron
and proton despite their different charges.  An important role in understanding
forces which bind multiple neutrons and protons ({\it nucleons}) into nuclei is
played by the {\it pion}, coupling to nucleons in an isospin-invariant way. With
exchange of pions and other heavier mesons, it was possible to understand the
masses of nuclei, with the deuteron (a neutron-proton bound state) a case in
point.  To the degree that nucleons in nuclei retain much of their identity,
one may think of nuclei as the first ``molecules'' of elementary particles.

\subsection{Quark Model}
\label{sec:quarkmodel}

Starting in the late 1940s, initially in cosmic rays but by 1953 also in
particle accelerators, a new degree of freedom, known as {\it strangeness},
began to be recognized in mesons and baryons \cite{GellMann:1953zza,*Nakano:1953zz}.  
Mesons and baryons could be classified into isospin multiplets
with their charges $Q$, the third component $I_3$ of their isospin $I$, and
their {\it hypercharge} $Y$ (a quantum number conserved in their strong
production) related by $Q = I_3 + Y/2$.  (The hypercharge is related to a
quantum number $S$, for ``strangeness,'' by $Y = S + B$, where $B$ is baryon
number.) However, in the 1950s it was not yet understood why certain isospin
multiplets appeared and not others, and how the observed ones were related to
one another.

By the early 1960s, it became clear that low-lying baryons included the
nucleon isospin doublet $(n,p)$ with $Y=1$, an isospin singlet $\Lambda$ and
an isospin triplet $\Sigma^-,\Sigma^0,\Sigma^+$ with $Y=0$, and an isospin
doublet $\Xi^-,\Xi^0$ with $Y=-1$.  These could be unified into an
eight-dimensional representation of the symmetry group SU(3)
\cite{GellMann:1961ky,*Neeman:1961jhl}.  The lowest-lying mesons, including
the pion and charged and neutral {\it kaons}, also could be fit into an
eight-fold multiplet along with a predicted meson called the $\eta$, soon
discovered \cite{Pevsner:1961pa}.

Given the spin $J=1/2$ and parity $P=+$ of the neutron and proton, their
partners in the SU(3) octet were predicted (and eventually observed) to have
$J^P = 1/2^+$.  But by the early 1960s a multiplet of resonant particles
with $J^P = 3/2^+$ was also taking shape:  an isoquartet $\Delta^-,\Delta^0,%
\Delta^+,\Delta^{++}$ with $Y=1$, a heavier isotriplet $\Sigma^{*-},%
\Sigma^{*0},\Sigma^{*+}$ with $Y=0$, and a still heavier isodoublet $\Xi^{*-},
\Xi^{*0}$ with $Y=-1$.  The SU(3) scheme predicted that these were members of
a ten-dimensional representation, to be completed by a predicted isosinglet
$\Omega^-$.  It also predicted an equal-spacing rule $M(\Omega) - M(\Xi^*)
= M(\Xi^*) - M(\Sigma^*) = M(\Sigma^*) - M(\Delta)$.  The second equality was
known to hold, and the predicted $\Omega$ was discovered in 1964, cementing
confidence in SU(3) \cite{Barnes:1964pd}.

The quark model \cite{GellMann:1964nj,*Zweig:1981pd} (see also
\cite{Petermann:1965qlk}) provided an explanation of SU(3), with the quarks
forming a fundamental triplet out of which all SU(3) representations
could be built.  For example, the ten-dimensional baryon representation
containing four $\Delta$, three $\Sigma^*$, two $\Xi^*$, and one $\Omega$
could be regarded as the totally symmetric $qqq$ combinations of $u$, $d$, and
$s$.  The dynamics of quarks featured prominently at the 1966 International
Conference on High energy Physics in Berkeley.  It seemed possible to describe
several hundred resonant particles in terms of three quarks for baryons and
a quark-antiquark pair for mesons.

A nagging question dealt with quark statistics.  The $\Delta^{++}$ was seen
as a ground state of three $u$ quarks with a total $J=3/2$, implying total
symmetry in its space $\times$ spin wave function.  But as a state of fermions,
its total wave function should be {\it antisymmetric}.  The invention of
another degree of freedom \cite{Greenberg:1964pe,*Han:1965pf}, now called {\it
 color}, in which every $qqq$ state could be totally antisymmetric, solved this
problem, and provided a basis for the interaction of quarks with one another
through the exchange of {\it gluons}.  This picture came to be known as {\it
quantum chromodynamics}, or QCD.

QCD also explained why quarks could form only integrally-charged states,
with their fractional charges masked by binding to other pairs of quarks
or antiquarks.  However, the question remained, to this day, why other
integrally-charged states such as $q q \bar q \bar q$ or $qqqq \bar q$, were
not observed.

Significant evidence for the reality of quarks came from deep inelastic
scattering of electrons on protons at the Stanford Linear Accelerator Center
\cite{Bloom:1969kc,*Breidenbach:1969kd}, recoiling against pointlike objects
consistent with quarks.  That these objects indeed appeared to have fractional
charge was indicated by a comparison of deep inelastic electron scattering with
that of neutrinos (see, e.g., \cite{Perkins:1972zt}).

The light quarks $u,d,s$ were eventually joined by heavier ones: $c$ (``charm")
\cite{Aubert:1974js,*Augustin:1974xw}, $b$ (``beauty'' or ``bottom'')
\cite{Herb:1977ek,*Ueno:1978vr}, and $t$ (``top'') 
\cite{Abe:1995hr,*Abachi:1995iq}.  In
contrast to the light quarks, whose properties and effective masses inside
mesons and baryons were strongly affected by the QCD interaction, $c$ and $b$
are amenable to approximately nonrelativistic descriptions, as their masses
($\sim 1.5$ and $5$ GeV, respectively) exceed their typical kinetic energies
(a few hundred MeV) inside mesons and baryons.  (Top quarks form only a
fleeting association with other quarks before they decay weakly, having a mass
of more than twice that of the $W$ boson.)

\subsection{Quantum Chromodynamics}
\label{sec:qcd}

The theory of the strong interactions, QCD, was born in a mathematical
investigation by Yang and Mills \cite{Yang:1954ek} of isotopic spin as a
{\it gauge theory}.  In contrast to electrodynamics, the quanta of a gauged
isospin theory carry charges.  One consequence of this is a different scale
dependence of the interaction strength.  The electrodynamic force becomes
stronger at short distances (large momentum-transfer scales), while in a
Yang-Mills type of theory the force becomes weaker at short distances
(``asymptotic freedom''). The relevant calculation, not then understood as
signaling asymptotic freedom, first appeared in a theory based on gauged SU(2)
symmetry \cite{Khriplovich:1969aa}. Asymptotic freedom was noticed by 't Hooft
in 1972 but never published \cite{tHooft:1998qmr};
and calculated in all generality for Yang-Mills
({\it non-Abelian gauge}) theories by Gross and Wilczek 
\cite{Gross:1973id,*Gross:1973ju,*Gross:1974cs} and Politzer \cite{Politzer:1973fx,*Politzer:1974fr}.

A gauged SU(3) as the theory of the strong interactions contains the
``color'' ingredient necessary to understand why the ground states of baryons
have quark wave functions that are {\it symmetric} in space $\times$ spin
$\times$ {\it flavor}, where the last term denotes the quark label $u,d,s,
\ldots$.  Each quark comes in one of three colors, and a wave function totally
antisymmetric in color can be constructed by taking one of each color.

The behavior of the strong interaction at {\it long} distances is also 
different from that of the electromagnetic interaction.  The gluons, quanta
of the strong interactions, interact with one another in such a way that
lines of force between a quark and an antiquark bunch up into a tube of
essentially constant cross-section area, leading by Gauss' Law to a constant
force at large distances, or a linearly rising potential.  (For recent comments
on this picture see Refs.\ \cite{Brambilla:2014jmp,*Shepherd:2016dni} and
references therein.)  When this potential
becomes strong enough, a new quark-antiquark pair is created, shielding the
color charges of the original pair.  Thus quark confinement is an essential
consequence of QCD.

The strengthening of the QCD coupling at low momentum scales and large distance
scales means that perturbation theory is unsuitable in that regime.  The
leading method for dealing with this behavior is to put spacetime onto a
{\it lattice}.  The limit as the lattice spacing $a$ tends to zero then may
be taken.  However, the presence of pions, with a very long Compton wavelength,
means that the total spatial extent of the lattice has to be large.  Coupled
with the need to take $a \to 0$, this leads to the requirement of very large
lattices, and typically fictional pions which are somewhat heavier than the
real ones.

The time-dependence of a spatial-lattice state can be described by taking
Euclidean time, whereby a dependence $e^{-imt}$ in Minkowski space is converted
to $e^{-m\tau}$, where $\tau \equiv it$.  In the limit $\tau \to \infty$,
a matrix element will behave as $e^{-m_0 \tau}$, where $m_0$ is the mass of
the lightest contributing intermediate state.  Subtracting off this
contribution, one can obtain, with some sacrifice in accuracy, the contribution
of the next-lowest intermediate state, and this process can be repeated until
statistical limitations set in.

Lattice QCD has been very successful in reproducing the masses of known states
involving $u,d,s,c,b$ quarks.  It has also been the leading means for
calculating {\it form factors} and {\it pseudoscalar meson decay constants},
which are more sensitive to wave function details.  This is particularly so now
that virtual light-quark-antiquark pairs have been taken into account in the
{\it unquenched} approximation.  The most sophisticated calculations even
consider virtual $c \bar c$ pairs when calculating properties of states
containing $b$ quarks.  Remaining possible sources of uncertainty include the
need for proper treatment of chiral fermions and the use of chiral perturbation
theory for extrapolation of calculations involving pions down to their
physical mass.

\subsection{QCD Motivated Models}
\label{sec:models}

\subsubsection{Potential Models}

The discoveries of the charmed and beauty quarks, and their rich $c \bar c$ and
$b \bar b$ spectra, led to approximate descriptions of their spectra by
nonrelativistic potential models 
\cite{Appelquist:1974zd,*Appelquist:1974yr,*Eichten:1974af,*Eichten:1978tg,%
Eichten:1979ms,*Quigg:1979vr,*Grosse:1979xm},
including those with relativistic corrections 
\cite{Godfrey:1985xj,*Barnes:2005pb}.  
The short-distance behavior of the interquark potential could
be described by a Coulomb-like potential, suitably modified by a logarithmic
correction due to asymptotic freedom, while the long-distance behavior was
linear in the separation $r$ (see the above description of quark confinement).
An interpolation between these two behaviors was provided by a potential
logarithmic in $r$ \cite{Quigg:1977dd}, for which the spacing between
$Q \bar Q$ levels was independent of the mass $m_Q$ of the quark $Q$, as is
nearly the case for $c \bar c$ and $b \bar b$ systems.

Treating light quarks in bound states as having effective masses of several
hundred MeV, and taking into account spin-spin (hyperfine) interactions among
them, it was even found possible to bypass many details of potential models,
gaining an insight into masses of light mesons and baryons or those containing
no more than one heavy quark ($c$ or $b$).  This approach was pioneered in
Ref.\ \cite{DeRujula:1975qlm} and applied, for example, to baryons containing
$b$ quarks in Refs.~\cite{Karliner:2006ny,*Karliner:2008sv}.

\subsubsection{Diquarks}

A baryon is made of three color-triplet quarks, coupled up to a color singlet
using the antisymmetric tensor $\epsilon_{\alpha \beta \gamma}$, where the
indices range from 1 to 3.  Each quark pair must then act as a color
antitriplet.  Under some circumstances it is then useful to consider a baryon
as a bound state of a color triplet quark and a color-antisymmetric
antitriplet {\it diquark}.  The color antisymmetry of the diquark requires
its space $\times$ spin $\times$ flavor wave function to be {\it symmetric},
where {\it flavor} denotes quark identity ($u,d,s,\ldots$).  For example,
the $u$ and $d$ quarks in the isosinglet baryon $\Lambda$ are in an S wave
(space symmetric) and an isospin zero state (flavor antisymmetric), so they
must be in a spin zero state (spin antisymmetric).  The spin of the $\Lambda$
is then carried entirely by the strange quark, consistent with its measured
magnetic moment \cite{Olive:2016xmw}.

Some light-quark resonances have been identified as candidates for
diquark-antidiquark bound states \cite{Jaffe:1976ig,*Jaffe:1976ih,Jaffe:1977cv},
with the last noting a relation to baryon-antibaryon resonances
\cite{Rosner:1968si,*Dalkarov:1970qb} reminiscent of the original Fermi-Yang
model of the pion \cite{Fermi:1959sa} as a nucleon-antinucleon bound state.
The past light-quark pentaquark candidates brought attention to a role 
diquarks can play in formation of such systems 
\cite{Karliner:2003dt,*Jaffe:2003sg}. Even though these candidates did 
not survive experimental scrutiny (see Sec 2.2), the discussion on the 
role of diquarks in shaping the structure of ordinary and exotic baryons 
\cite{Jaffe:2004ph,*Selem:2006nd} is very much alive today.

\subsubsection{Tightly Bound Multiquark States}

In addition to the above light-quark resonances, some authors have postulated
that new resonances including one or more heavy quarks are candidates for
tightly bound diquark-antidiquark states \cite{Maiani:2004vq,*Maiani:2014aja,%
*Brodsky:2014xia,*Chen:2015dig,Maiani:2015vwa,Maiani:2016wlq}. Thus, the $X(3872)$
first observed decaying to $J/\psi \pi^+ \pi^-$ \cite{Choi:2003ue} would be
interpreted as a bound state of a $cu$ diquark and a $\bar c \bar u$
antidiquark.  We shall discuss the merits and drawbacks of this assignment
presently.

\subsubsection{Hadrocharmonium}

The resonance $X(3872)$ mentioned above can be regarded as a charmonium state
embedded in light hadronic matter, called {\it hadrocharmonium} in Ref.\
\cite{Dubynskiy:2008mq}.  This classification is motivated by the observation
that multiquark states including a $c \bar c$ pair appear to contain only a
single charmonium state, whereas one might expect the wave function to involve
a linear combination of several charmonium states in a hadronic molecule or
generic multiquark state.

\subsubsection{Molecular States}

The wave functions of many exotic multiquark states such as $X(3872)$ appear
to consist, at least in part, of pairs of hadrons each containing one heavy
quark.  Thus, one can identify $X(3872)$ as a bound or nearly bound state of
$(D^0 = c \bar u)(\bar D^{*0} = \bar c u)$ + (charge conjugate), as we shall
discuss in Sec.\ \ref{sec:quarkoniumlike}.  Such assignments are favored if the
constituents can be bound via exchange of a light pseudoscalar, such as
pion \cite{Tornqvist:1993ng,*Tornqvist:2003na,Tornqvist:2004qy,Karliner:2015ina} or possibly $\eta$ 
\cite{Karliner:2016ith}.  As in the case of the deuteron, pion exchange is not
the whole story, but, where permitted, dominates the long-range force.

\subsubsection{Cusps and Anomalous Triangle Singularities}

When a decay process involves three particles in the final state, the
proximity of S-wave thresholds in two-body rescattering can lead to behavior
which can mimic a resonance while only consisting of a cusp.  Kinematic
enhancements can also be due to {\it anomalous triangle singularities} (for an
early manifestation in pion-nucleon scattering see \cite{Peierls:1961zz}),
in which resonance-like behavior is seen when all participants in rescattering
approach the mass shell.  Triangle singularities and methods to identify true
resonances as $S$-matrix poles have been recently discussed in Refs.\
\cite{Guo:2014iya,*Szczepaniak:2015eza,*Liu:2015taa,Mikhasenko:2015vca,%
Guo:2016bkl}.

\section{LIGHT MULTIQUARK CANDIDATES}
\label{sec:lighthadrons}

\subsection{Light Meson Multiquark Candidates}
\label{sec:lightmesons}

The P-wave $q \bar q$ states of the three light quarks $q=u,d,s$ consist of
$^3P_{0,1,2}$ and $^1P_1$ nonets with positive parity.  Here the superscript
denotes the quark-spin multiplicity $2S_q+1$, while the subscript denotes the
total spin $J$.  The $J=0$ states can couple to a pair of pseudoscalar mesons
in an S wave, and hence their widths and masses are strongly influenced by
these couplings.  Indeed, one can regard them as linear combinations of
$q \bar q$ and meson-meson states.  The latter can be thought of as $qq \bar q
\bar q$, or {\it tetraquarks}.  A systematic classification of light $J=0$
mesons as tetraquarks was made by Jaffe \cite{Jaffe:1976ig,*Jaffe:1976ih,%
Jaffe:1977cv}. 

The two-pseudoscalar-meson channel strongly affects the production and decay
of the nonstrange $J=0$ mesons $f_0(980)$ (isoscalar) and $a_0(980)$
(isovector) \cite{Olive:2016xmw}.  They lie very close to the $K \bar K$
threshold and thus may be thought of, in part, as either $K \bar K$ bound
states or tetraquarks containing an $s \bar s$ pair.  The
$f_0$ is seen to decay predominantly to $\pi \pi$, but is produced primarily
in processes which provide an initial $s \bar s$ pair, such as $B_s^0 \to
J/\psi~f_0$ \cite{Stone:2013eaa}.

Another light-quark system in which meson, rather than quark, degrees of
freedom play an important role is the $f_1$ or $a_1$ system decaying to $K \bar
K \pi$ with mass around 1420 MeV.  The Dalitz plot near this total mass shows
$a_0$ or $f_0$, $K^*$,$\bar K^0$ resonances between each final-state pair
\cite{Rosner:1987zk}.  The $f_1(1420)$ thus may not be a genuine resonance
but rather a kinematic effect known as a {\it triangle singularity}
\cite{Debastiani:2016xgg}.

Cusp-like behavior in scattering amplitudes near S-wave thresholds for new
final states is widespread \cite{Rosner:2006vc}.  For one example,
diffractive photoproduction of $3 \pi^+ 3 \pi^-$  exhibits a dip near
$p \bar p$ threshold \cite{Frabetti:2001ah,*Frabetti:2003pw}.  There may also
be a $p \bar p$ resonance or bound state near this mass, but the question is
not settled \cite{Rosner:2006vc}.

\subsection{Light Baryon Multiquark Candidates}
\label{sec:lightbaryons}

The quark model for baryons has been very successful in describing them as
$qqq$ states, including those with nonzero internal orbital angular momentum.
However, final meson-baryon states (and thus states of $q \bar q + qqq$) play
an important role as well.  A case in point is the resonance known as
$\Lambda(1405)$, with $J^P = 1/2^-$. It has a history going as far back as the
late 1950s \cite{Dalitz:1959dn,*Dalitz:1959dq,*Dalitz:1960du}; for a recent
understanding of its structure see Ref.~\cite{Meissner:2015pdg}.
Its nature is
still being debated, though it is a reasonable candidate for a $\bar K N$ bound
state.  The quark model predicts three $J^P = 1/2^-$ isospin-zero baryons: a
flavor singlet with quark spin 1/2 and two flavor octets, one with quark spin
1/2 and the other with quark spin 3/2.  The $\Lambda(1405)$ appears to be
mainly the flavor singlet, with smaller admixtures of the two octets
\cite{Isgur:1978xj}.  Two other states, $\Lambda(1670,1/2^-)$ and $\Lambda%
(1800,1/2^-)$ \cite{Olive:2016xmw}, are the orthogonal mixtures.  Couplings to
the channels $\Sigma \pi$, $N \bar K$, and $\Lambda \eta$ probably play some
role in the mixing \cite{Isgur:1978wd,*Koniuk:1979vy}.
\smallskip

A candidate for a $K^+n$ resonance called $\Theta^+(1540)$, whose minimal quark
content would be $\bar s uudd$, was observed in the early 2000s
\cite{Nakano:2003qx, *Stepanyan:2003qr, *Kubarovsky:2003fi}.  However, it was not
confirmed in further experiments \cite{Danilov:2008zza}
and appears to have been a kinematic effect \cite{Rosner:2003ia}.

\section{HEAVY-LIGHT MULTIQUARK CANDIDATES} 
\label{sec:heavylighthadrons}

\subsection{Heavy-Light Meson Multiquark Candidates}
\label{sec:heavylightmesons}

The S-wave states of a charmed quark and a light ($u,d,s$) antiquark are the
pseudoscalar mesons $D^0$, $D^+$, and $D_s^+$ ($^1S_0$) and the vector mesons
$D^{*0}$, $D^{*+}$, and $D_s^{*+}$ ($^3S_1$).  The P-wave states naturally
divide into those with light-quark total angular momentum $j = 1/2$ ($J^P_j =
0^+_{1/2},~1^+_{1/2}$) and $j=3/2$ ($J^P_j = 1^+_{3/2},2^+_{3/2}$)
\cite{DeRujula:1976ugc}.  They are predicted to decay to ground-state charmed
mesons in the following ways, where $P$ stands for $\pi$ or $K$:
$0^+_{1/2} \to DP~(L=0)$; $1^+_{1/2} \to D^*P~(L=0)$;
$1^+_{3/2} \to D^*P~(L=2)$; $2^+_{3/2} \to (D,D^*)P~(L=2)$.
The states with $j=3/2$ decaying via D-waves are expected to be narrow, and
indeed correspond to the observed $D_1(2420)$, $D_2(2460)$, $D_{s1}(2536)$,
and $D_{s2}(2573)$ \cite{Olive:2016xmw}.  (Here the subscript denotes total
$J$.)  Information is fragmentary on the nonstrange $j=1/2$ states but there
exists a broad candidate for the nonstrange $0^+_{1/2}$ state with mass
$M=2318 \pm 29$ MeV and width $\Gamma = 267 \pm 40$ MeV \cite{Olive:2016xmw}.
When both strange and nonstrange candidates for the same $(J,j)$ are seen, the
strange candidate is about 115 MeV heavier than the nonstrange candidate.  Thus
we would expect a strange $0^+_{1/2}$ state around $115 + 2318 = 2433$ MeV,
above the $D K$ threshold of 2362 MeV.

What came as a surprise was the observation by the BaBar Collaboration
\cite{Aubert:2003fg} of a candidate for the strange $0^+_{1/2}$ state at 2317
MeV, more than 100 MeV below na\"{\i}ve expectations and 45 MeV below $DK$ 
threshold. It was seen instead to decay to $D_s \pi^0$ via an isospin-violating
transition.  A hint of a strange state at 2460 MeV, decaying to $D_s \pi^0
\gamma$, was also seen.  Its confirmation \cite{Besson:2003cp,Krokovny:2003zq,%
Aubert:2003pe} supplied a candidate for the strange $1^+_{1/2}$ state
\cite{Aubert:2004pw,Choi:2015lpc}, 40 MeV below $D^* K$ threshold.

Proposals for explaining the displacement of $D_{s0}(2317)$ and $D_{s1}(2460)$
masses from their expected values included the formation of $D^{(*)} K$
molecules or bound states \cite{Barnes:2003dj}, the existence of tetraquarks
\cite{Dmitrasinovic:2005gc,Cheng:2003kg,Terasaki:2003qa,Bracco:2005kt}, and a
realization of chiral symmetry which predicted the observed mass pattern 
a number of years earlier 
\cite{Nowak:1992um,Bardeen:1993ae,Bardeen:2003kt,Nowak:2003ra}.
The yet-to-be-detected conjectured bottom analogues of
$D_{s0}(2317)$ $D_{s1}(2460)$ are discussed in Sec.~\ref{sec:beyonddetected}.

\subsection{Heavy-Light Baryon Multiquark Candidates}
\label{sec:heavylightbaryons}

Threshold effects can involve heavy mesons and light-quark baryons,
or heavy baryons and light-quark mesons.  An example of the former is a
charmed baryon resonance $\Lambda_c(2940)$, seen decaying to $D^0 p$
\cite{Aubert:2006sp}.  The mass was seen to be just below $D^{*0}p$
threshold, suggesting a bound state or molecular interpretation
\cite{Ortega:2012cx,*Zhang:2012jk,*Zhang:2014ska,*Dong:2014ksa,Xie:2015zga}.
Recently the LHCb Collaboration \cite{Aaij:2017vbw} has analyzed the $D^0 p$
amplitude in $\Lambda_b \to D^0 p \pi^-$ and finds a resonance favoring $J^P
= 3/2^-$ at a mass of $2944.8^{+3.5}_{-2.5} \pm 0.4 ^{+0.1}_{-4.6}$ MeV with a
width of $27.7^{+8.0}_{-6.8}\pm 0.8^{+5.2}_{-10.4}$ MeV.  The $J^P$ assignment
is consistent with an S-wave state of a $D^{*0}$ and a proton.

Following the alleged discovery of the $\Theta(1540)$ pentaquark candidate (see
Sec.\ 2.2) a $\bar c uudd$ state was claimed by the H1 Collaboration at HERA in
Hamburg \cite{Aktas:2004qf},
corresponding to an effective mass of 3.1 GeV in the $D^{*\pm} p^{\mp}$ system.
It was not confirmed with further data \cite{Karshon:2009ws}.

\section{HEAVY QUARKONIUM-LIKE MULTIQUARK CANDIDATES}
\label{sec:quarkoniumlike}

\subsection{Ground rules}
\label{sec:groundrules}

In this section we shall discuss states containing two heavy quarks $Q$ which
cannot be represented as simple $Q \bar Q$ excitations, but which require some
admixture of light quarks as well.  The notation $X$ will stand for neutral
``cryptoexotic'' states with likely $Q \bar Q q \bar q$ content. States in this
category with $J^{PC} = 1^{--}$ which can couple directly to a virtual photon
will be denoted $Y$, while those with a charged light-quark pair (e.g., $u \bar
d$) will be denoted $Z_c$ (when the heavy pair is $c \bar c$) or $Z_b$ (when
the heavy pair is $b \bar b$).  Finally, $P_c$ or $P_b$ will denote a state
such as $c \bar c uud$ or $b \bar b uud$ (``pentaquark''). 

The spectrum of $X$, $Y$, $Z$ states is particularly rich for charmonium.
Some controversy exists over the quark content, spin, and parity of many of
these states.  A useful reference to the experimental literature is
contained in Ref.~\cite{Eidelman:2016}.
We shall not 
discuss in any detail states which we believe to have conventional $Q \bar Q$
assignments, concentrating instead on $X,~Y,~Z,$ and $P_Q$ candidates.

\subsection{The $X(3872)$ State}
\label{sec:x3872}

\begin{figure}[htbp]
\hbox{\hskip\lfskip
\includegraphics[width=1.01\typewidth]{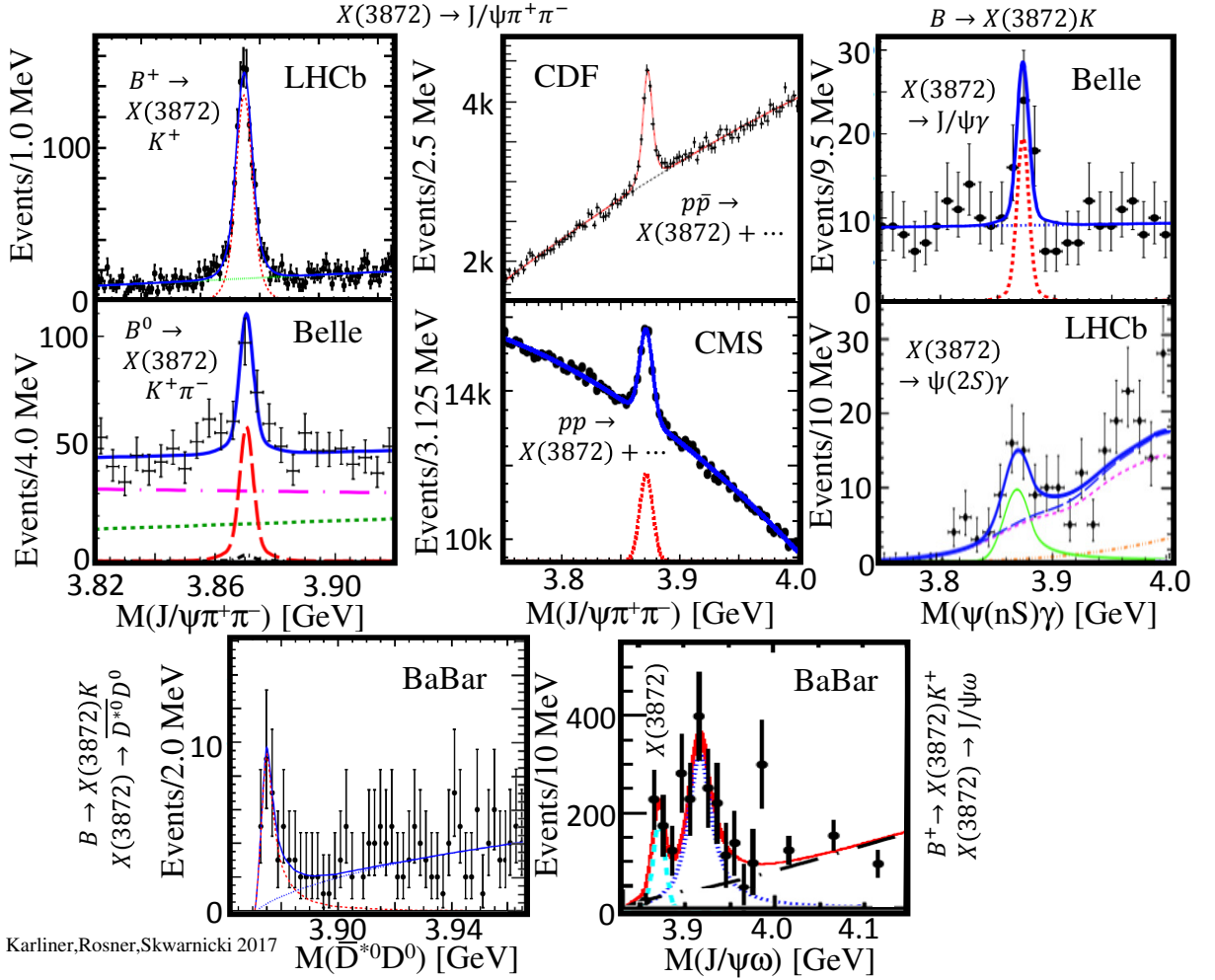}
}
\caption{Production and decay of the $X(3872)$ state. 
Detailed figure descriptions can be found in the original references, 
from which the plots have been adapted:
{\bf top row} 
left Ref.~\cite{Aaij:2015eva},
middle Ref.~\cite{Aaltonen:2009vj},
right \cite{Bhardwaj:2011dj},
{\bf middle row}
left Ref.~\cite{Bala:2015wep},
middle Ref.~\cite{Chatrchyan:2013cld},
right Ref.~\cite{Aaij:2014ala},
{\bf bottom row}
left Ref.~\cite{Aubert:2007rva},
and
right Ref.~\cite{delAmoSanchez:2010jr}.
}
\label{fig:x3872}
\end{figure}

The first evidence for a multiquark state involving $c \bar c$ and light quarks
came from the decay $B \to K \pi^+ \pi^- J/\psi$, in which the $\pi^+ \pi^-
J/\psi$ system showed a narrow peak around 3872 MeV \cite{Choi:2003ue}.  It
has been confirmed by many other experiments \cite{Olive:2016xmw}, as illustrated 
in Fig.~\ref{fig:x3872}.   
Its width is less than 1.2 MeV, and its $J^{PC}$ has been established as $1^{++}$
\cite{Aaij:2015eva}.

The mass of $X(3872)$, whose 2016 average \cite{Olive:2016xmw} is $3871.69 \pm
0.17$ MeV, is sufficiently close to the threshold for $D^0 \bar D^{*0}$,
namely $(1864.83 \pm 0.05) + (2006.85 \pm 0.05) = (3871.68 \pm 0.07)$ MeV,
that one cannot tell whether it is a candidate for a bound state or resonance
of $D^0 \bar D^{*0}$.  Clearly, however,
the neutral-$D$ channel must play an important role in the makeup of $X(3872)$,
as also evidenced by a large fall-apart rate of the $X(3872)$ to $D^0 \bar D^{*0}$, 
once the kinematic threshold is 
exceeded \cite{Gokhroo:2006bt,Aubert:2007rva,Adachi:2008sua} (Fig.~\ref{fig:x3872}).
The $D^+ D^{*-}$ threshold, $(1869.59 \pm 0.09) + (2010.26 \pm 0.05) = (3879.85
\pm 0.10)$ MeV, is sufficiently far from $M(X(3872))$ that the charged-$D$
channel appears to play a much less important role in its composition.

The quark makeup of $X(3872)$ thus should include an important $c \bar c u \bar
u$ component.  Confirmation of this point is provided by the observation of
both $X(3872) \to \omega J/\psi$ ($\omega\to\pi^+\pi^-\pi^0$) \cite{Abe:2005ax,delAmoSanchez:2010jr}
and $X(3872) \to \rho^0 J/\psi$ ($\rho^0\to\pi^+\pi^-$) \cite{Abulencia:2005zc,Aaij:2015eva}, implying that
the $X(3872)$ is a mixture of isospins zero and one \cite{Tornqvist:2004qy} (Fig.~\ref{fig:x3872}).
There also appears to be a $c \bar c$ $\chi_{c1}(2P)$ component to the
$X(3872)$ wave function, as indicated by the ratio of the radiative decays 
to $\gamma \jpsi$ and $\gamma \psi(2S)$ \cite{Aaij:2014ala} (Fig.~\ref{fig:x3872}),
\begin{equation}
\label{Rpsigamma:eq}
R_\gamma\equiv \frac{{\cal B}(X(3872) \to \psi(2S) \gamma)}
{{\cal B}(X(3872) \to \jpsi \gamma)} =2.46 \pm 0.64 \pm 0.29\,.
\end{equation}
The measured value is consistent with pure charmonium and a mixture of
charmonium and a molecular state, but not with a pure molecular state.
Additional rather robust evidence for the $\bar c c$ component is provided by
the relatively large cross section for prompt production of $X(3872)$ in
$p\bar p$ \cite{Acosta:2003zx, Abazov:2004kp} and~$pp$ collisions
\cite{Aaij:2011sn,Chatrchyan:2013cld,Aaboud:2016vzw} (Fig.~\ref{fig:x3872}), 
closely following behavior of the $\psi(2S)$ state.

In particular, Ref.~\cite{Esposito:2015fsa} uses ALICE data on the
production of light nuclei with $p_T \lsim 10$ GeV in Pb-Pb collisions at
$\sqrt{s_{NN}} =2.76$ TeV to estimate the expected production cross
sections of such nuclei in $pp$ collisions at high $p_T$.  Hypertriton,
helium-3, and deuteron production cross sections are compared to the CMS
results for prompt production of $X(3872)$ \cite{Chatrchyan:2013cld}.  Fig.~1
of Ref.~\cite{Esposito:2015fsa} shows that the latter is orders of magnitude
larger than the former.  Also the dependence of the prompt production of
$X(3872)$ on its transverse momentum and pseudo-rapidity, as well as the ratio
of the prompt production to the production in $B$ meson decays, closely follow
those of the $\psi(2S)$ charmonium state, pointing to the same production
mechanism \cite{Abazov:2004kp,Aaboud:2016vzw}. 

The cross section for prompt production of these light nuclei falls
rapidly with $p_T$ because  they are rather large. As soon as $p_T$ is
bigger than their inverse radius, the probability of forming such weakly
bound molecular states becomes very small.  The $X(3872)$ binding energy
is much smaller than the 2.2 MeV deuteron binding energy. Therefore the
spatial extent of the molecular component must be much bigger than the
already large deuteron size.

We can estimate the inverse size of the
$X(3872)$ molecular component using the formula
\begin{equation}
\label{X872size:eq}
1/r = \sqrt{2 \mu |\Delta E|}
\end{equation}
where $\mu=967$ MeV  is the reduced mass and $\Delta E$ is the binding
energy.  $\Delta E$ is at most 0.2 MeV, probably less. This gives $1/r
\lsim 20$ MeV, corresponding to radius of $\gsim 10$ fermi, really
huge. With such a large radius the cross section for production of the
molecular component at $p_T \gsim 10$ GeV is expected to be negligible.
Therefore $X(3872)$ must have a significant $c \bar c$ component, whose
size is the typical hadronic radius $< 1$ fermi, much smaller than the
size of the molecular component.\footnote{It was argued that the short-range
structure of the molecular wave function is difficult to predict
\cite{Guo:2014taa,Albaladejo:2017blx}, so large values of 
$R_\gamma$ and of prompt production cross-section are not incompatible 
with the molecular behavior of the wave function at large distances. 
This, however, does not imply that these experimental observations are 
natural expectations in the molecular model. We side with the argument 
that an admixture of charmonium $2^3P_1$ state offers the most natural 
explanation, and in fact, is not incompatible with the molecular 
behavior of $X(3872)$ at large distances.}

This of course raises the interesting question of how the mixing works for two
states whose sizes differ by at least a factor of 10.  Perhaps $X(3872)$ lives
long enough to make even a small spatial overlap sufficient for significant
mixing.\footnote{Alex Bondar, private communication.}
It is also possible that the molecular component occurs dynamically 
when the compact $X(3872)$ attempts to disintegrate to $D^0\bar D^{*0}$.

Hadronic molecules were proposed some time ago \cite{Tornqvist:1993ng,%
Voloshin:1976ap,DeRujula:1976zlg,Tornqvist:1991ks}.  One-pion exchange plays
an important (though not exclusive) role in facilitating binding.  The
attractive force between two states of isospin $I_{1,2}$ and spin $S_{1,2}$
transforms as
\begin{equation}
V \sim \pm I_1 \cdot I_2~S_1 \cdot S_2~~~{\rm for}~(qq,~q \bar q)~
{\rm interactions},
\end{equation}
and is expected to bind not only $D^0 \bar D^{*0} + {\rm c.c.}$ but many other
systems as well, including meson-meson, meson-baryon and baryon-baryon
\cite{Karliner:2015ina}.  In particular, there should be an analogue $X_b$ of
the $X(3872)$, near $B \bar B^*$ threshold ($10604.8 \pm 0.4$ MeV for neutral
$B$-s and $10604.5 \pm 0.4$ MeV for $B^+ B^{-*}$) \cite{Karliner:2014lta}.
Because the thresholds for charged and neutral pairs are so similar, isospin
impurity in the $X_b$ is expected to be small, and it should be mostly
isoscalar.

CMS and ATLAS have searched for the decay $X_b \to \Upsilon(1S)\pi^+
\pi^-$ \cite{Chatrchyan:2013mea,*Aad:2014ama}. The search in this particular
channel was motivated by the seemingly analogous decay $X(3872) \to \jpsi
\pi^+ \pi^-$. This analogy is misguided, however, because for an
isoscalar with $J^{PC} = 1^{++}$ such a decay is forbidden by $G$-parity
conservation \cite{MKtoCMS,*Guo:2014sca}.
Thus the null result of these searches does not tell us anything about 
the existence of $X_b$.

The bottomonium state $\chi_{b1}(3P)$ has been recently observed
\cite{Aad:2011ih,*Aaij:2014caa}.  The $X_b$ state could mix with it and share
its decay channels, just as $X(3872)$ is likely a mixture of a $\bar D D^*$
molecule and $\chi_{c1}(2P)$
\cite{Karliner:2014lta}. However, the mass difference between the observed
$\chi_{b1}(3P)$ state and $B\bar B^*$ thresholds is about 93 MeV, which
makes a significant mixing unlikely. In fact, the observed $\chi_{b1}(3P)$
mass is in excellent agreement with the potential model predictions made
over 20 years before its first observation \cite{Kwong:1988ae}, while mixing
would have likely affected its mass.

\subsection{Other Near-threshold Quarkonium-like Mesons}
\label{sec:thresholdstates}

\begin{figure}[htbp]
\hbox{\hskip\lfskip
\includegraphics[width=1.01\typewidth]{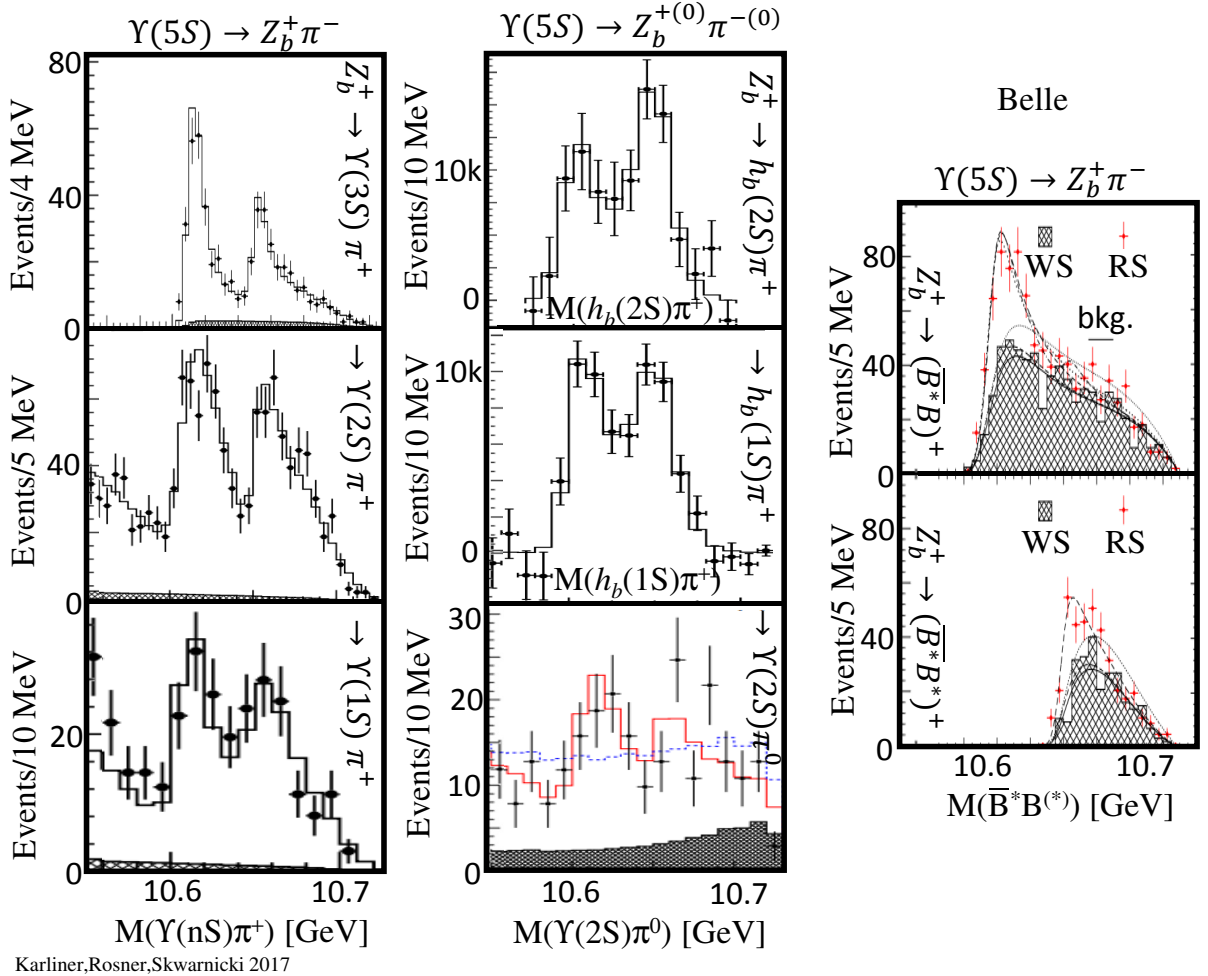}}
\caption{Observations of the $Z_b(10610)$ and $Z_b(10650)$ states.
Detailed figure descriptions can be found in the original references, 
from which the plots have been adapted:
{\bf left column} 
Ref.~\cite{Belle:2011aa},
{\bf middle column}
top and middle Ref.~\cite{Belle:2011aa},
bottom Ref.~\cite{Krokovny:2013mgx},
and {\bf right column}
Ref.~\cite{Garmash:2015rfd}.
}
\label{fig:zb}
\end{figure}

The cross sections for $e^+ e^- \to \Upsilon(1S,2S,3S)\pi^+\pi^-$ and $e^+ e^-
\to \Upsilon(1S) K^+ K^-$ were found to be surprisingly large near the peak of
the $\Upsilon(5S)$ resonance at $\sqrt{s} \sim 10.87$ GeV \cite{Abe:2007tk}.
One possible explanation of this enhancement was the existence 
of intermediate $b \bar b q_1 \bar q_2$ states ($q_i$ denotes a light quark)
decaying to $\Upsilon(nS) \pi$ or $\Upsilon(1S) K$
\cite{Karliner:2008rc}.  Unusual enhancements were also seen in the cross
sections for $e^+ e^- \to h_b(nP)\pi^+\pi^-~(n=1,2)$, where $h_b(nP)$ denotes
a spin-singlet $b \bar b$ resonance with radial quantum number $n$, orbital
angular momentum $L=1$, and total spin $J=1$ \cite{Adachi:2011ji}.  These
effects were found to be due to two charged bottomonium-like resonances in
$\Upsilon(5S)$ decays \cite{Belle:2011aa}.  The $\Upsilon(nS) \pi~(n=1,2,3)$
spectra are shown in Fig.~\ref{fig:zb}.  The peaks have been named $Z_b(10610)$ and
$Z_b(10650)$.  Similar peaks are seen in $M(h_b(nP)\pi^+\pi^-)~(n=1,2)$.
The review of Ref.\ \cite{Eidelman:2016} quotes the average masses as
$M(Z_b(10610)) = 10607.2 \pm 2.0$ MeV and $M(Z_b(10650) = 10652.2 \pm 1.5$
MeV.

The masses of the two peaks are very close to the thresholds for $B \bar B^*$
and $B^* \bar B^*$:  $(10604.0 \pm 0.3)$ MeV and $(10649.3 \pm 0.5$ MeV,
respectively.  This suggests that their wave functions should largely consist
of the respective S-wave ``molecular'' components $B \bar B^*$ and $B^* \bar
B^*$ \cite{Bondar:2011ev}.  
In fact, the $Z_b(10610)$ fall-apart rate to $B \bar B^*$ is large,
while there is no evidence for $Z_b(10650)\to B \bar B^*$, which prefers to 
decay to $B^* \bar B^*$ in spite of the smaller 
phase-space \cite{Garmash:2015rfd} (Fig.~\ref{fig:zb}).
The absence of similar effects just above the $B\bar B$ 
threshold of $(10558.6 \pm 0.3)$ MeV points to an important role of
one-pion exchange in the formation of these ``molecules'', as a pion cannot
couple to the pair of pseudoscalar mesons $B \bar B$ \cite{Karliner:2015ina}.
 
\begin{figure}[htbp]
\hbox{\hskip\lfskip\quad\hskip0.1\typewidth
\includegraphics[width=1.01\typewidth]{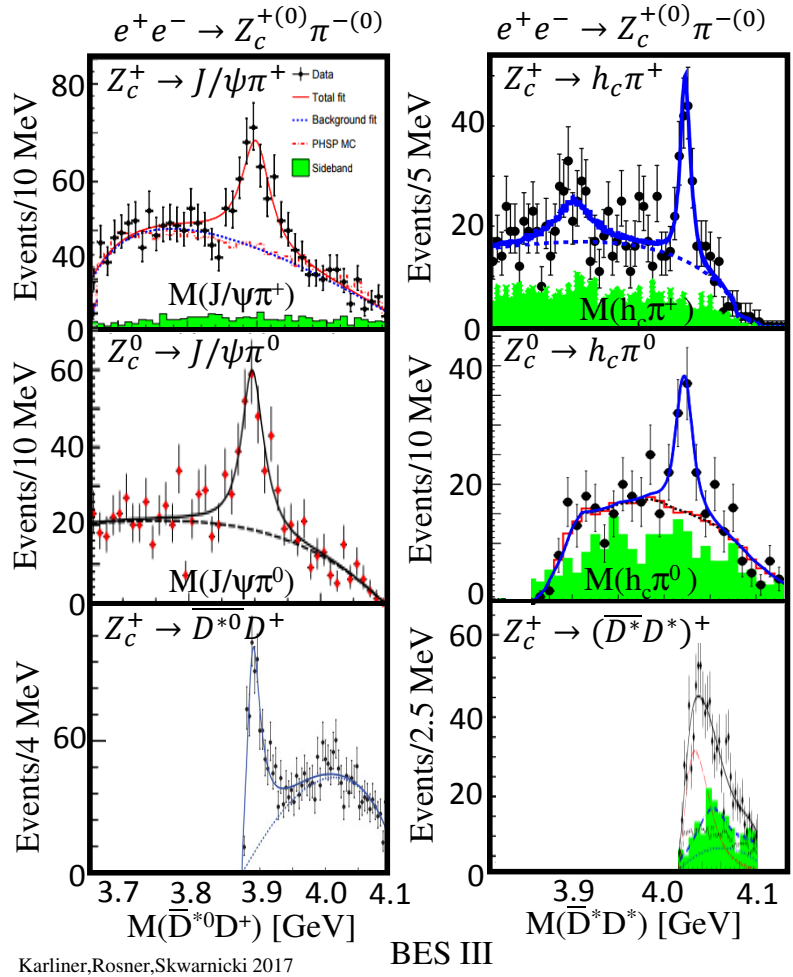}
\hskip-0.1\typewidth\quad}
\caption{Observations of the $Z_c(3900)$ and $Z_c(4020)$ states.
Detailed figure descriptions can be found in the original references, 
from which the plots have been adapted:
{\bf left column} 
top Ref.~\cite{Ablikim:2013mio},
middle Ref.~\cite{Ablikim:2015tbp},
bottom Ref.~\cite{Ablikim:2013xfr},
{\bf right column}
top Ref.~\cite{Ablikim:2013wzq},
middle Ref.~\cite{Ablikim:2014dxl},
and bottom Ref.~\cite{Ablikim:2013emm}.
}
\label{fig:zc}
\end{figure}

A counterpart to the $Z_b$ system has been observed in exotic charmonium
states.  The thresholds for [neutral, charged] $D \bar D^*$ pairs are [$(3871.7
\pm 0.1),~(3879.8 \pm 0.1)$] MeV, while the thresholds for [neutral, charged]
$D^* \bar D^*$ pairs are [$(4013.7 \pm 0.1),~(4020.52 \pm 0.1)$] MeV.  States
near both these thresholds, respectively called $Z_c(3900)$ and $Z_c(4020)$,
have been observed in decays of the vector meson candidate $Y(4260)$ (see next
section).  The $Z_c(3900)$ is seen in the $\pi \pi J/\psi$ final state as a
peak in $M(\pi J/\psi)$ \cite{Ablikim:2013mio,Liu:2013dau,Xiao:2013iha,%
Ablikim:2015tbp} and in the $\pi D \bar D^*$ final state as a peak in
$M(D \bar D^*$ \cite{Ablikim:2015gda,Ablikim:2013xfr}, as illustrated in
Fig.~\ref{fig:zc}.  
Its averaged mass is quoted as $(3891.2 \pm 3.3)$ MeV \cite{Eidelman:2016}.  
The $Z_c(4020)$ is
seen in the $\pi \pi h_c$ final state as a peak in $M(\pi h_c)$
\cite{Ablikim:2013wzq,Ablikim:2014dxl} and in the $\pi D^* \bar D^*$ final
state as a peak in $M(D^* \bar D^*)$ \cite{Ablikim:2013emm,Ablikim:2015vvn}
(Fig.~\ref{fig:zc}).
Its averaged mass is quoted as $(4022.9 \pm 2.8)$ MeV \cite{Eidelman:2016}.
As in the case of the exotic bottomonium $Z_c$ states, the absence of $D \bar
D$ peaks is circumstantial evidence in favor of a role for pion exchange in
forming molecules of open-flavor pairs.  As mentioned earlier, such molecules
were anticipated shortly after the discovery of charm \cite{Voloshin:1976ap}.

As mentioned in Sec.~\ref{sec:models}, many explanations of these near-threshold $Z_b$
and $Z_c$ states abound.  The close correlation between peaks and thresholds
would have to be regarded as a coincidence in potential models.  The grouping
of multiple quarks in an exotic hadron depends to some extent on their masses;
an example is the predominance of $QQ$ color antitriplet diquarks in $QQ\bar q_1 \bar q_2$
hadrons due to the tight binding of the heavy quarks $Q$ with one another.
Explanations based on genuine tetraquarks require the observation of isospin
partners; the $Z_c(3900)$ may be the charged partner of $X(3872)$, though this
interpretation is complicated by the isospin impurity of the latter.  Many of
the vector states to be described in the next section may admit of a
hadrocharmonium explanation \cite{Dubynskiy:2008mq}.  Finally, experience with
light-quark systems such as $f_0(980)$ and $\Lambda(1405)$ indicates that
resonant vs.\ cusp behavior may be difficult to sort out when new channels
are opening up.  Experience with Feshbach resonances 
\cite{Feshbach:1958nx,*Feshbach:1962ut}, 
also associated with the opening of new channels, may be of
help here.

\subsection{Anomalous Vector States}
\label{sec:vectorstates}

The direct coupling of quarkonium states with $J^{PC} = 1^{--}$ to virtual
photons has made them particularly easy to observe.  The charm and bottom
quarks were first observed as a result of these couplings in the (S-wave)
$1^3S_1$ states $J/\psi(1S) = c \bar c$ and $\Upsilon(1S) = b \bar b$,
respectively. Weaker couplings to virtual photons also are possessed by the
(D-wave) $1^3D_1$ states.  Here we use the notation $n^{2S+1}L_J$, where $n$
is the radial quantum number, $S$ denotes quark spin, $L$ is represented by $S,
P, D, F, \ldots$ for $L=0,1,2,3,\ldots,$ and $J$ denotes total spin of the
state.  Candidates for such ``conventional'' vector quarkonia include the
following, where we use the name assigned in Ref.\ \cite{Olive:2016xmw} and
give the approximate mass in MeV:

\noindent
Charmonium: $J/\psi(1S)(3097),\psi(2S)(3686),\psi(1D)(3770),\psi(3S)(4040),
\psi(2D)(4160)$,\\ $\psi(4S)(4415)$;

\noindent
Bottomonium: $\Upsilon(1S)(9460),\Upsilon(2S)(10023),\Upsilon(3S)(10355),
\Upsilon(4S)(10579),\Upsilon(5S)(10860)$,\\ $\Upsilon(6S)(11020)$.

\begin{figure}[htbp]
\hbox{\hskip\lfskip
\includegraphics[width=1.01\typewidth]{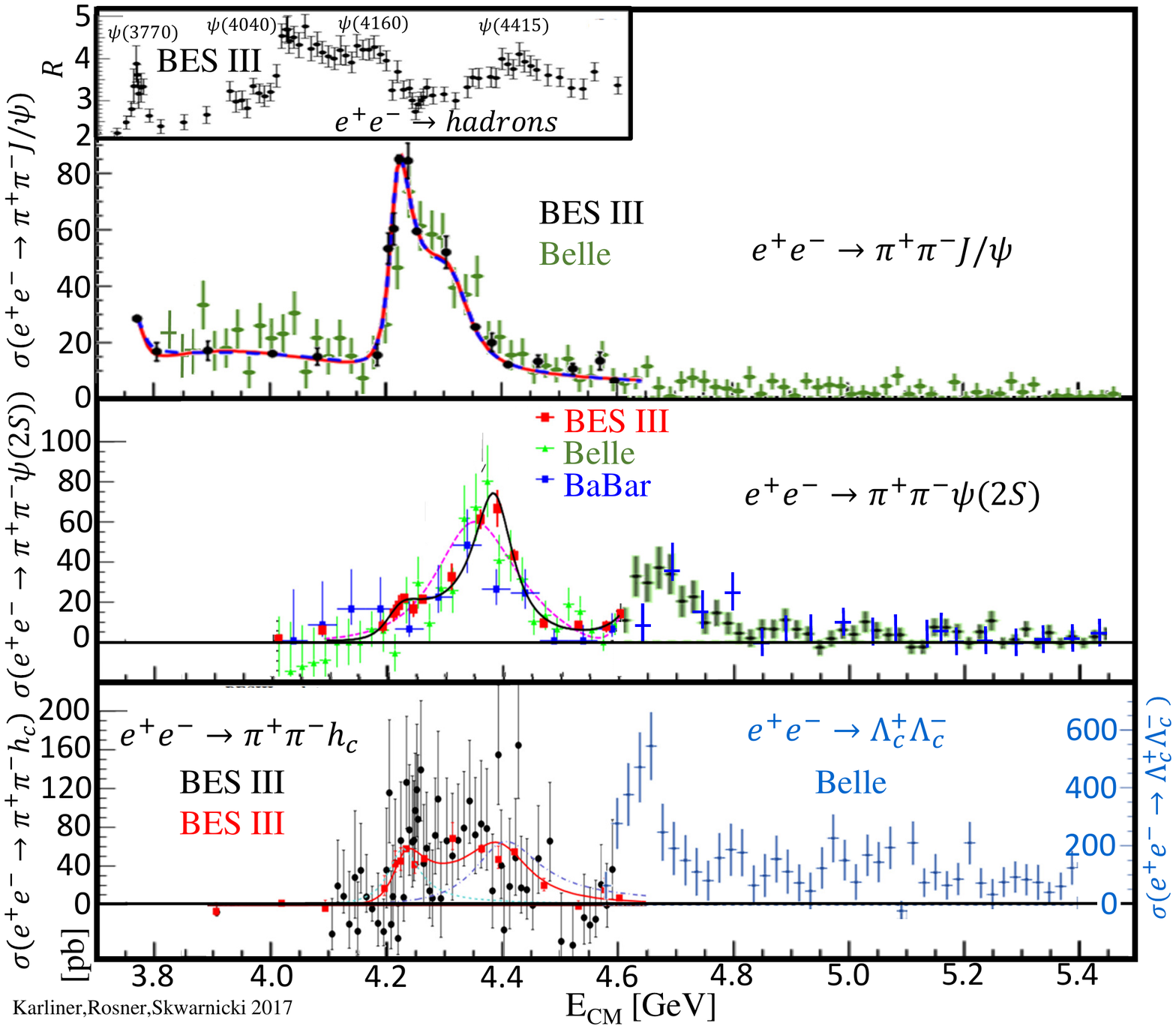}}
\caption{Measurements of cross-sections for $e^+e^-$ annihilation:
(top left inset)
to hadrons expressed in units of $R$ \cite{Hu:2017lqp},
(top row)
to $\pi^+\pi^-\jpsi$ \cite{Ablikim:2016qzw,Liu:2013dau},
(middle row)
to $\pi^+\pi^-\psi(2S)$ \cite{Ablikim:2017oaf,Lees:2012pv,Wang:2014hta},
(bottom row, left)
to $\pi^+\pi^-h_c$ 
(black/red points are from two different energy scans by BESIII) 
\cite{BESIII:2016adj}, and
(bottom row, right)
to $\Lambda_c^+\Lambda_c^-$ \cite{Pakhlova:2008vn}.
The displayed curves were fitted to the BESIII data  
(see Refs.~\cite{Ablikim:2016qzw,Ablikim:2017oaf,BESIII:2016adj}).
}
\label{fig:Y}
\end{figure}

The ratio $R \equiv \sigma(e^+ e^- \to {\rm hadrons})/\sigma(e^+ e^- \to \mu^+
\mu^-)$ as measured by BESIII \cite{Hu:2017lqp} (Fig.~\ref{fig:Y})
peaks prominently around 4040 MeV, and noticeably
just above 4400 MeV, motivating the charmonium 3S and 4S assignments for these
peaks.  A peak associated with the 2D candidate is less prominent, as befits a
D-wave state whose coupling to a virtual photon is suppressed.
A prominent feature of $R$ is a steep drop around $E_{\rm
c.m.} = 4.2$ GeV.  The change in $R$ is more than one unit, which could
signify the total suppression of charm production ($\Delta R = -4/3$).
Such a sharp dip is often associated with the opening of a new S-wave channel
\cite{Rosner:2006vc}, as in the case of $I=0$ $\pi \pi$ scattering near
$K \bar K$ threshold. Indeed, the lowest-lying two-body S-wave state into which
a $c \bar c$ pair can fragment is $D \bar D_1-{\rm c.c.}$ \cite{Close:2005iz},
where $D_1$ is a P-wave bound state of a charmed quark and a light ($\bar u$
or $\bar d$) antiquark with $J^P = 1^+$.  The minus sign corresponds to the
negative $C$ eigenvalue.  The lightest established candidate for $D_1$ has a
mass of about 2.42 GeV/$c^2$, corresponding to a threshold of 4.285 GeV.

The cross sections $\sigma(e^+ e^-\to f)$, where $f$ are specific final states,
differ considerably from one another (see the mini-review by Eidelman {\it et
al.} \cite{Eidelman:2016}).  For this reason, we briefly describe the apparent
resonant activity in each final state.  Just as in light-quark spectroscopy,
mixing of quark-model configurations can lead to eigenstates favoring 
individual channels.

\underline{$\pi \pi J/\psi$ {\it final state:}}
The cross section for $e^+ e^- \to \pi^+ \pi^- J/\psi$, as first seen in the
radiative return process $e^+ e^- \to \gamma \pi^+ \pi^- J/\psi$ by the BaBar
Collaboration \cite{Aubert:2005rm} and confirmed in several other experiments
\cite{Eidelman:2016}, shows a prominent peak around 4260 MeV.  It could be a
$D \bar D_1$ state with about 25 MeV of binding energy \cite{Ding:2008gr}.  
Weaker evidence for
a $J^{PC} = 1^{--}$ state around 4008 MeV presented by the Belle Collaboration
\cite{Yuan:2007sj} has not been confirmed by others.  Recently the BESIII
Collaboration has reported two new structures in $\sigma(e^+ e^-\to \pi^+ \pi^-
J/\psi)$:  one with a mass of $(4222.0 \pm 3.4)$ MeV and a broader one
with a mass of $(4320\pm13)$ MeV \cite{Ablikim:2016qzw} (Fig.~\ref{fig:Y}).
The first could be identified with a shifted $Y(4260)$, while the second has
been proposed as an artifact of interference among $\psi(4160)$, $\psi(4415)$,
and nonresonant background \cite{Chen:2017uof}.
The lower $Y(4260)$ mass, with an asymmetric high-mass shoulder, 
was previously proposed based on the older data in the 
$D\bar D_1$ molecular model \cite{Cleven:2013mka}. 

\underline{$\pi \pi \psi(2S)$ {\it final state:}}
Peaks in the effective mass of $\pi^+ \pi^- \psi(2S)$ states have been seen
by Belle and BaBar around 4360 and 4660 MeV 
\cite{Aubert:2007zz,Lees:2012pv,Wang:2007ea,Wang:2014hta} (Fig.~\ref{fig:Y}).  
The former (``$Y(4360)$") is roughly in the mass range expected
for a charmonium 4S state, so the true 4S $c \bar c$ amplitude might be shared
between the $Y(4360)$ and the $\psi(4415)$, with the rest of the $\psi(4415)$
wavefunction as a shallowly bound S-wave state of $D_1~(J^P  = 1^+)$ and
$D^*~(J^P = 1^-)$.  Alternatively, 4360 MeV is a plausible threshold for
production of a $D(0^+) \bar D^*(1^-)$ pair.  The latter (``$Y(4660)$") could be
associated with a peak at a nearby mass, about 4630 MeV, in $e^+ e^-
\to \Lambda_c^+ \Lambda_c^-$ \cite{Pakhlova:2008vn} (Fig.~\ref{fig:Y}).
The most precise data on the $\pi^+\pi^-\psi(2S)$ channel in the 
lower-peak region was recently published by BESIII (Fig.~\ref{fig:Y}), 
which found a significant evidence for a state at $4.21$ GeV, 
perhaps the same one as observed in the $\pi^+\pi^-\jpsi$ channel, 
and improved the $Y(4360)$ mass determination to $4384\pm4$ MeV 
\cite{Ablikim:2017oaf}.

\underline{$\pi \pi h_c$ {\it final state:}}
The $h_c(3525)$ is the lowest-lying ($n=1$) $n^1P_1$ charmonium level, first
seen by the CLEO Collaboration \cite{Rosner:2005ry}.  It is curious that, as a
spin-singlet level, it should have been produced in $e^+ e^- \to \pi^+ \pi^-
h_c$, as first observed by the CLEO Collaboration \cite{CLEO:2011aa} and
reported recently by the BESIII Collaboration \cite{BESIII:2016adj}.
Normally one would expect a virtual photon to produce a $c \bar c$ spin-triplet
state, so the process must be violating heavy-quark symmetry, perhaps via an
intermediate open-flavor-pair state \cite{Bondar:2011ev} in which the
correlation between heavy-quark spins is lost.
Two resonant structures are seen, one at $4218\pm5$ MeV and
a broader one at $4392\pm7$ MeV (Fig.~\ref{fig:Y}).  
The first is consistent
with the BESIII observation in the $\pi^+ \pi^- J/\psi$ and $\pi^+ \pi^- \psi(2S)$ 
final states mentioned
above, while the second could be an artifact of interference among
$\psi(4160)$, $\psi(4415)$, and nonresonant background \cite{Chen:2017uof}.

\underline{{\it Open charm final states:}}
A comprehensive analysis of the behavior of the cross section for production
of open charm final states has been made in Ref.\ \cite{Xue:2017xpu}.  The
analysis supports the identification, mentioned above, of the $Y(4260)$ as
mainly a molecular state of $D \bar D_1(2420)$.  The resonance line shapes
for $e^+ e^- \to D^* \bar D^*$ and $e^+ e^- \to D_s^* \bar D_s^*$ can
be satisfactorily explained with contributions from $\psi(4040)$, $\psi(4160)$,
and $\psi(4415)$, assuming suitable relative phases. 

\underline{{\it Resonances $\Upsilon(5S)$ and $\Upsilon(6S)$:}}
The behavior of $R$ above $B \bar B$ threshold exhibits two bumps, called
$\Upsilon(5S)$ and $\Upsilon(6S)$, with respective masses $10890 \pm 3$ and
$10993^{+10}_{-3}$ MeV \cite{Eidelman:2016}.  An example of the shape of
these bumps is given in Ref.\ \cite{Santel:2015qga}.  Decay modes common to
both include $B_{(s)}^{(*)} \bar B_{(s)}^{(*)}$, $\pi \pi \Upsilon(1S,2S,3S)$,
and $\pi \pi h_b(1P,2P)$.  As in the case of $Y(4260) \to \pi \pi h_c$, the
latter class of decays violates heavy-quark symmetry and points to the
role of open-flavor intermediate states \cite{Bondar:2011ev}.  The large
decay widths for the transitions to $\pi \pi \Upsilon(nS)$ may be understood
as enhancements of decay rates due to the intermediate states $Z_b(10610)$
and $Z_b(10650)$ \cite{Karliner:2008rc}.
Several other decay modes, including $f_0 \Upsilon(1S)$, $\eta\Upsilon(1S,2S)$,
and $\pi^+ \pi^- \Upsilon(1D)$, are reported for $\Upsilon(5S)$.

\subsection{Other Exotic Meson Candidates Detected in $B$ Decays}
\label{sec:inbdecays}

The first evidence for an explicitly exotic charged $Q\bar{Q}u\bar{d}$ state,
$Z_c(4430)^{+}\to\psi(2S)\pi^{+}$, was claimed by the Belle collaboration in
$\bar{B}\to\psi(2S)\pi^+ K$ decays ($K=K^0_S$ or $K^-$) in 2007
\cite{Choi:2007wga}, well before the other charged candidates were observed in
$e^+e^-\to\pi^{\mp} Z_{b,c}^{\pm}$ (see Sec.~\ref{sec:thresholdstates}).
This state has a vivid experimental history. It was first claimed as a narrow
peak ($\Gamma=45^{+35}_{-18}$ MeV) 
in the invariant $\psi(2S)\pi^{\pm}$ mass distribution \cite{Choi:2007wga}, 
with parameters obtained by a na\"{\i}ve fit to this distribution with ad hoc
assumptions about the shape of the background from excited kaons, $K^{*}\to
K\pi^{+}$, dominating such $B$ decays. This observation was soon questioned by
the BaBar experiment  \cite{Aubert:2008aa}.  In response, the Belle experiment
published amplitude analyses with a realistic model of $K^*$ resonances, first
performed on the Dalitz plane~\cite{Mizuk:2009da}, later also including angular
information from $\psi(2S)\to\ell^+\ell^-$ decays \cite{Chilikin:2013tch},
which pointed to a significant  $J^P=1^+$ $Z_c(4430)^{+}\to\psi(2S)\pi^{+}$
contribution, albeit much broader than initially claimed
($\Gamma=200^{+49}_{-58}$ MeV).  Later, the LHCb collaboration confirmed the
Belle results in a similar amplitude analysis performed on a much larger sample
of $B$ decays \cite{Aaij:2014jqa}, and demonstrated consistency of the
$Z_c(4430)^{+}$ peak with a resonant hypothesis using an Argand diagram. They
also demonstrated a need for other significant contributions than $K^{*0}\to
K^-\pi^{+}$ to $\bar{B}^0\to\psi(2S)\pi^+ K^-$ decays without any assumptions
about $K^{*}$ resonances, other than limiting their spin in the relevant low
$K^-\pi^{+}$ mass region \cite{Aaij:2015zxa}.

The Belle collaboration claimed to have spotted $Z_c(4430)^{+}\to\jpsi\pi^+$
in $\bar{B}^0\to\jpsi \pi^+K^-$ decays, this time producing a dip in the
$\psi(2S)\pi^+$ mass distribution via interference with an even broader
($\Gamma=370^{+\phantom{0}99}_{-149}$~MeV) second $1^+$ state, 
$Z_c(4200)^+\to\jpsi\pi^+$, \cite{Chilikin:2014bkk}.  There was also some
indication for a second $Z_c^+\to\psi(2S)\pi^+$ state around that mass with
$0^-$ or $1^+$ quantum numbers in the LHCb data \cite{Aaij:2014jqa}.

The Belle collaboration also reported evidence for charged $\chi_{c1}\pi^+$
resonances, the $Z_c(4050)^+$ and $Z_c(4250)^+$, in the amplitude analysis of
$\bar{B}^0\to\chi_{c1}\pi^+ K^-$ decays, but could not determine their quantum
numbers~\cite{Mizuk:2008me}.  BaBar saw an enhancement in the same $\chi_{c1}
\pi^+$ mass region, but suggested it could be a reflection of $K^*$ resonances
\cite{Lees:2011ik}.  Without an amplitude analysis, their results do not
contradict the Belle results. 

As broad states, the charged $Z_c^+$ candidates are poor candidates for
molecules of $D$ and $\bar{D}$ excitations.  They have not been reported in
prompt production at the Tevatron or LHC, thus also making poor candidates for
tightly bound tetraquark states.  It is remarkable that they have not been
observed in the $e^-e^+\to\pi^{\mp} Z_{c}^{\pm}$ reaction; and, vice versa,
the $Z_c^+$ states observed there have not been seen in $B$ decays. 
This points to hadron-level forces responsible for these structures, 
perhaps via hadron rescattering in $B$ decays, as such forces are expected to
be sensitive to details of production mechanisms.  Future high-statistics
amplitude analyses of $B$ decays in the upgraded LHCb and Belle experiments
should shed more light on these effects.

The history of the $X(4140)$ state has some parallels to the $Z_c(4430)^+$ saga.
It was first claimed in 2008 as a narrow peak ($\Gamma=11.7\,^{+9.1}_{-6.2}$
MeV) observed by the CDF collaboration in the invariant $\jpsi\phi$ mass
distribution from $B^+\to\jpsi\phi K^+$ decays \cite{Aaltonen:2009tz}.  The
existence of such a narrow, near-threshold state was questioned by the LHCb
experiment \cite{Aaij:2012pz}. 
The CMS experiment confirmed its existence, however, with somewhat larger width
\cite{Chatrchyan:2013dma}.  Later the LHCb experiment analyzed the biggest to
date sample of $B^+\to\jpsi\phi K^+$ decays, and performed the first amplitude
analysis of this channel, thus providing more realistic subtraction of
the $B^+\to\jpsi K^{*+}$, $K^{*+}\to\phi K^+$ backgrounds
\cite{Aaij:2016iza,Aaij:2016nsc}.  The LHCb data are consistent with a
near-threshold $\jpsi\phi$ resonance, however, with a much broader width
($\Gamma=83\,^{+30}_{-25}$~MeV) than initially measured.  LHCb determined its
quantum numbers to be $J^{PC}=1^{++}$. 

Since the 2011 update of the CDF analysis, there was a hint for a second
$X(4274)\to\jpsi\phi$ state in the same $B^+$ decay mode \cite{Aaltonen:2011at}.
A second $\jpsi\phi$ mass enhancement was visible in the CMS data, but at a
higher mass \cite{Chatrchyan:2013dma}.  The amplitude analysis by LHCb
confirmed the $X(4274)$ state with high statistical significance and determined
its quantum numbers to be also $1^{++}$ \cite{Aaij:2016iza,Aaij:2016nsc}.  Two
$0^+$ states at higher masses, $X(4500)$ and $X(4700)$, were also needed for
a good description of the LHCb data. 

The D0 experiment presented an evidence for prompt production of $X(4140)$ in $p\bar{p}$ 
collisions at Tevatron \cite{Abazov:2015sxa}. It is puzzling why the $X(4140)$ 
width observed in this inclusive measurements was narrow ($\Gamma=16\pm13$ MeV) and why the 
$X(4274)$, $X(4500)$ and $X(4700)$ were not observed. This observation awaits a confirmation.

The Belle experiment, which was lacking statistics in the $B^+\to\jpsi\phi K^+$
channel, looked for $\jpsi\phi$ states in $\gamma\gamma$ collisions.  They
obtained evidence for a narrow $X(4350)$ state ($\Gamma=13^{+18}_{-10}$~MeV)
and saw no other $\jpsi\phi$ 
mass peaks \cite{Shen:2009vs}. The $X(4350)$ state awaits confirmation as well.

The origin of the $\jpsi\phi$ states, among which the $X(4140)$ and $X(4274)$
should be considered experimentally established, is far from clear.  Their
masses do not fall into the mass intervals near the pairs of excitations of
the $D_s$ ($\bar{D}_s$) mesons with {\bf matching quantum numbers} for S-wave
interactions, bound by $\eta$ exchange \cite{Karliner:2016ith}.  Explanation of
the $X(4140)$ as related to a $D_s^{\pm}D_s^{*\mp}$ cusp
\cite{Swanson:2015bsa,Aaij:2016nsc}, relies on broadening this threshold effect
via a poorly justified form factor.  Tightly bound tetraquark models can
account for a doublet of $1^{++}$ states only in an approach in which ``good''
(color antitriplet) and ``bad'' (color sextet) 
diquarks are allowed \cite{Stancu:2009ka}.
In the tetraquark model using only ``good'' diquarks, it was suggested that $X(4274)$ is not a $1^{++}$ state
but a superposition of two states with $0^{++}$ and $2^{++}$ \cite{Maiani:2016wlq}.
However, such components of $X(4274)$ are disfavored by more than $7$ 
standard deviations by the LHCb analysis (Table 7 in Ref.~\cite{Aaij:2016nsc}).
It was suggested that $X(4274)$, $X(4500)$ and $X(4700)$ states are conventional 
$3^3P_1$, $4^3P_0$ and $5^3P_0$ charmonium states, respectively \cite{Ortega:2016hde}.
However, no explanation of why $4^3P_1$ and $5^3P_1$ states 
would not be also visible in the $\jpsi\phi$ decay mode was offered.
None of the $X\to\jpsi\phi$ states observed in $B^+\to\jpsi\phi K^+$ decays
is seen in the $\jpsi\omega$ decay mode probed in $B^+\to\jpsi\omega K^+$ decays
(see Fig.\ 43 in Ref.~\cite{Olsen:2017bmm}), suggesting 
that the $s\bar s$ pair is among the constituents of these states.
With no plausible theoretical interpretation of all four of them together, 
they may have different origins or be some complicated 
artifacts of rescaterring of $D_{s(J)}^{(*)}$ meson-antimeson pairs.
Future higher-statistics samples of $B^+\to\jpsi\phi K^+$ decays 
may allow probing the nature of these structures in a less model-dependent way
and shedding more light on their nature.

A near-threshold enhancement in the $\jpsi\omega$ mass distribution in
$B\to\jpsi\omega K$ decays was first reported by Belle \cite{Abe:2004zs}. 
BaBar later resolved this structure into two mass peaks, identified with 
$X(3872)\to\jpsi\omega$ decay (Sec.~\ref{sec:x3872}) and the state at
$3919\pm4$ MeV with rather narrow width, $\Gamma=31\pm11$ MeV
\cite{delAmoSanchez:2010jr} (Fig.~\ref{fig:x3872}).  
Both Belle and BaBar observed a state at similar
mass and width in $\gamma\gamma$ collisions \cite{Uehara:2009tx,*Lees:2012xs}.
It is commonly assumed, but not proven, that these mass structures are due to
the same state as the one called $X(3915)$, with $0^{++}$ or $2^{++}$ as likely
quantum numbers.  This state is too narrow to be a conventional charmonium
triplet $P$-state (for a full discussion see Ref.~\cite{Olsen:2017bmm}) at
masses where decays to $D\bar{D}^*$ and $D\bar{D}$ are allowed.  It was
recently proposed that mixing of the $2^3P_2$ charmonium state with a molecular
$D\bar{D}^*$ or $D^*\bar{D}^*$ component could be responsible for $X(3915)$
\cite{Ortega:2017qmg,*Baru:2017fgv}.

\subsection{Quarkonium-like Pentaquark Candidates}
\label{sec:pentaquarks}

A possibility of four quarks and one antiquark binding together
was anticipated from the beginnings of the quark 
model \cite{GellMann:1964nj,*Zweig:1981pd},
later reinforced by QCD, in which a diquark can effectively 
act as an antiquark, thus two diquarks and one antiquark can 
attract each other by the same means as three antiquarks do in an ordinary 
antibaryon.  However, even today, we can't directly predict from QCD 
if such bound states can live long enough to have any measurable effects.
Pentaquarks made only out of up and down quarks  
lack useful experimental signatures to distinguish them from
ordinary baryons. Pentaquarks with a flavored antiquark would 
decay strongly to a baryon and a flavored meson, a final state
which cannot be produced in a decay of an ordinary baryon. 
While some pentaquark candidates of that type were claimed in the past
experiments, none of them survived scrutiny 
of additional data 
(see Secs.~\ref{sec:lightbaryons},\ref{sec:heavylightbaryons}).

\begin{figure}[bhtp]
\begin{center}
\includegraphics[width=0.5\typewidth]{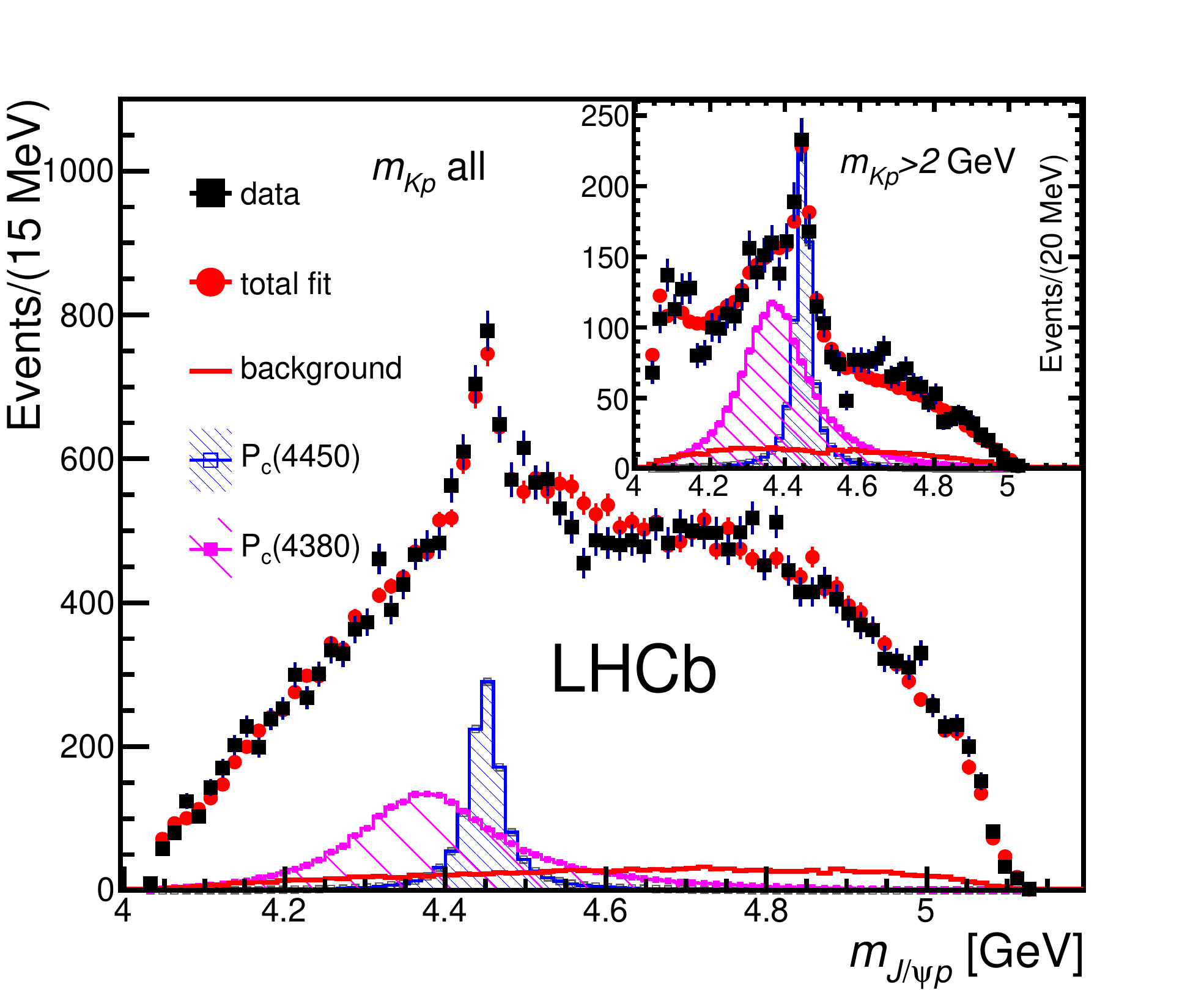}
\end{center}
\caption{Observation of the pentaquark candidates $P_c(4450)^+$ 
and $P_c(4380)^+$ decaying to $\jpsi p$ in the amplitude 
analysis of $\Lambda_b\to\jpsi p K^-$ decays by the
LHCb collaboration. Adapted from Ref.~\cite{Aaij:2015tga}.
}
\label{fig:Pc}
\end{figure}

In 2015, the LHCb experiment observed a rather narrow ($\Gamma\sim40$ MeV) 
structure in the $\jpsi p$ mass distribution in 
$\Lambda_b\to\jpsi p K^-$ decays \cite{Aaij:2015tga}, as shown
in Fig.~\ref{fig:Pc}.
Since the heavy $c\bar c$ pair in the $\jpsi$ cannot be created 
during hadronization with rates which would lead to such observation,
this structure makes for a convincing $uudc\bar{c}$ candidate.
Its statistical significance is much larger than any of the previous 
pentaquark candidates, thus this effect is not going to fade away with
additional data.  LHCb demonstrated at the 9$\sigma$ level
that the $\jpsi p$ mass peak cannot be due to 
reflections of excited $\Lambda$ states decaying to $pK^-$, 
with almost no assumptions about such $\Lambda^*$ baryons, 
which dominate this $\Lambda_b$ decay mode \cite{Aaij:2016phn}.  
An amplitude analysis of these data, which used 13 well established 
$\Lambda^*$ resonances as a model for the $pK^-$ component, revealed 
that two $\jpsi p$ resonances were needed for a reasonable 
description of data: the narrow $P_c(4450)^+$ ($\Gamma=39 \pm 20$ MeV) 
and the lighter and wider $P_c(4380)^+$ ($\Gamma=205 \pm 88$ MeV). 
Both states had a very high statistical significance ($12\sigma$ and $9\sigma$,
respectively), albeit depending on much stronger model assumptions
\cite{Aaij:2015tga}.  The Dalitz plot pattern of their intensities implies
they should have opposite parities.  The spin combinations involving $3/2$ and
$5/2$, in either order, were preferred.  In addition to pentaquarks with
$uudc\bar{c}$ quarks bound together in one confining
volume by color forces, also baryon-meson molecules bound 
by residual color forces, similar to those responsible for creation of nuclei,
can have the same quark content.  In fact, a $\Sigma_c\bar{D}^*$ molecular
state was predicted by Karliner and Rosner around the $P_c(4450)^+$ mass
\cite{Karliner:2015ina}.  This model requires $J^P=3/2^-$ and provides a
natural explanation for its narrow width.  The $p\chi_{c1}$ mass threshold
coincides with the $P_c(4450)^+$ mass \cite{Meissner:2015mza}.  Such a
molecular state, or cusp, would require $J^P=3/2^+$.  Molecular bound states or
cusps don't offer any explanation for the broad $P_c(4380)^+$ state,
nor can they lead to spin as high as $5/2$ in this mass range. Rescattering of
ordinary baryons and mesons, via the so-called triangle anomaly, must
happen in an S-wave to be pronounced, thus cannot account for spin $5/2$ either
\cite{Guo:2015umn,Liu:2015fea,Mikhasenko:2015vca}.  The tightly bound
pentaquark model can generate such high spin for $P_c(4450)^+$ 
via orbital angular momentum between quarks \cite{Maiani:2015vwa}
and can account for the wider $P_c(4380)^+$.  So far, the rich mass spectrum
necessarily resulting from such quark confinement has not been experimentally
observed.  It is also not clear why such a pentaquark state would be narrow,
with the large phase-space available for $\jpsi p$ decays and the spatial
proximity of $c$ and $\bar c$. 
It was suggested that momentary separation of $c$ and $\bar c$, followed by 
immediate hadronization, can be a result of a production mechanism pushing them
in opposite directions \cite{Lebed:2015tna}.  Such a cartoon model is lacking
predictive power, thus is difficult to confirm or dismiss.  The LHCb
Collaboration did not assign statistical or systematic significance to the 
determined quantum number preference. Therefore, it is premature to draw
strong conclusions about possible interpretations of the $P_c^+$ states based
on this preference.  It is more than likely that the LHCb model of the
$\Lambda^*$ states was incomplete, since about 60 $\Lambda^*$ states are
predicted in the relevant mass range by the quark model \cite{Aaij:2015tga}, 
and in fact some of them were observed in various analyses of the $KN$
scattering data, but were too model-dependent to earn labels of
well-established states by the PDG \cite{Olive:2016xmw}.  Coupled channels,
especially $(\Sigma\pi)_{I=0}$, are likely to make significant contributions
as well.  More $\Lambda_b\to\jpsi p K^-$ data are already available to LHCb.
It is hoped their improved amplitude analysis will shine more light on 
the nature of these $\jpsi p$ mass structures.

The LHCb analyzed also the Cabibbo-suppressed channel $\Lambda_b\to\jpsi p
\pi^-$ \cite{Aaij:2016ymb}.  With much fewer events, complications from many
known $p\pi^-$ resonances, and the possibility of an exotic contribution from
the $Z_c(4200)^-\to\jpsi\pi^-$ state, the results were inconclusive. 
The data are fully compatible with the $P_c(4380)^+$ and $P_c(4450)^+$ 
contributing to this final state at the expected level, but also 
compatible with no such contributions if the $Z_c(4200)^-$ is allowed.

There have been no claims of spotting the $P_c^+$ states in prompt production
at LHC, which would have favored a tightly bound pentaquark model.  Molecular
or tightly bound $\jpsi p$ states should be reachable in photoproduction at
JLab \cite{Wang:2015jsa,*Kubarovsky:2015aaa,*Karliner:2015voa,*Blin:2016dlf},
where several experimental searches for them are under way.

\section{BEYOND DETECTED STATES}

\label{sec:beyonddetected}

\smallskip
\underline{$QQ\bar{Q}\bar{Q}$:}

The question of whether there exist bound states of two heavy quarks $Q=(c,b)$
and antiquarks $\bar Q = (\bar c, \bar b)$, distinct from a pair of
quark-antiquark mesons, has been debated for more than forty years.
It has drawn substantial interest recently
\cite{Karliner:2016zzc,
*Bai:2016int,
*Wang:2017jtz,
*Richard:2017vry,
Anwar:2017toa,
Eichten:2017ual,
*Vega-Morales:2017pmm,
*Hughes:2017xie}.
Ref.~\cite{Karliner:2016zzc} predicted
$M(X_{cc\bar c \bar c}) = 6{,}192 \pm 25$ MeV
and
$M(X_{bb\bar b \bar b}) = 18{,}826 \pm 25$ MeV,
for the $J^{PC} = 0^{++}$ states involving charmed and bottom tetraquarks,
respectively. 
Earlier predictions vary over a big range, with large error bars, 
cf. Table VII in Ref.~\cite{Karliner:2016zzc}. A more recent compilation of 
predicted values of $M(X_{bb\bar b\bar b}) - 2M(\eta_b)$ 
appears in Table I of Ref.~\cite{Anwar:2017toa}.
The proximity of the predicted $X_{bb\bar b \bar b}$ mass 
to $2M(\eta_b) = 18,798\pm5$ MeV \cite{Olive:2016xmw}
and the size of the theoretical errors
suggests that $X_{bb\bar b \bar b}$ either decays strongly with a rather
narrow width, or it is below the $\eta_b\eta_b$ threshold, in which case
one expects final states of hadrons from pairs of intermediate gluons,
and of hadrons or leptons from pairs of intermediate virtual photons.
Experimental search for these states in the relevant mass
range is highly desirable.  
Searches in the four-lepton and $\ell^+ \ell^- B \bar B$ final states have
been performed at the LHC \cite{Aad:2015oqa,*Khachatryan:2017mnf}.  These
are
devoted to the search for the standard-model Higgs boson decaying into two
light pseudoscalars $a$, which then decay to such final states as $\mu^+
\mu^-,~\tau^+\tau^-,$ and $b \bar b$.  These are ideal samples for the
searches advocated here.
\smallskip

\underline{Bottom analogues of $D^*_{s0}(2317)$ and $D_{s1}(2460)$:}

These $B_{sJ}$ states are the yet-to-be-discovered $b$-quark analogues of the
very narrow $D_{sJ}$ states seen by BaBar, CLEO and Belle
\cite{Aubert:2003fg,Besson:2003cp,Krokovny:2003zq,Aubert:2003pe}
$D_{s0}(2317)$ with $J^P=0^+$ and $m[D_{s1}(2460)]$ with $J^P=1^+$,
conjectured to be the 
chiral partners of $D_s$, $J^P=0^-$ and $D_s^*$, $J^P=1^-$, respectively
\cite{Bardeen:2003kt,Nowak:2003ra}.
A strong hint toward this conjecture is supplied by almost equal splitting
between the states of opposite parity \cite{Olive:2016xmw}:
$m[D_{s0}(2317)]- m[D_s]{=}349.4 \pm 0.6$ MeV${\approx}$
$m[D_{s1}(2460)]-m[D_s^*] \kern-0.2em = 347.3\pm 0.7$ 
MeV${\approx}$ constituent mass of light quarks.
Assuming approximately the same splitting in the bottom sector,
one expects $B_{s0}$ at $\sim5717$ MeV with $J^P = 0^+$
and $B_{s1}$ at $\sim5765$ MeV with $J^P = 1^+$.
They are also predicted by a lattice calculation \cite{Lang:2015hza}.
These states are likely to be observed at LHCb and might also be accessible at
Belle II in $e^+ e^- \to B_{s0} \bar B_s^*$ and $e^+ e^- \to B_{s1} \bar B_s$
\cite{Karliner:2015tga}.
\smallskip

\underline{Stable $bb\bar u\bar d$ tetraquark:}

Recently LHCb discovered the first doubly-charmed baryon $\Xi_{cc}^{++} =
ccu$ at $3621.40 \pm 0.78$ MeV \cite{Aaij:2017ueg},
very close to the theoretical prediction $3627 \pm 12$ MeV in 
Ref.~\cite{Karliner:2014gca}.\footnote{We refer the reader 
to Refs.~\cite{Aaij:2017ueg} and \cite{Karliner:2014gca} for an extensive 
list of other predictions, most of which quote much greater uncertainties.}
In Ref.~\cite{Karliner:2017qjm} the same theoretical approach was used to
predict a doubly-bottom tetraquark \,$T(bb\bar
u\bar d)$\, with $J^P{=}1^+$ at $10,389\pm 12$ MeV,\, 215 MeV below the
$B^-\bar B^{*0}$ threshold and 170 MeV below threshold for decay to
$B^-\bar B^0 \gamma$.  Similar conclusions were obtained in
Refs.~\cite{Eichten:2017ffp,*Czarnecki:2017vco,*Francis:2016hui}.  The
$T(bb\bar u\bar d)$\, is therefore stable under strong and electromagnetic
(EM) interactions and can only decay weakly, the first exotic hadron with such
a property.  The predicted lifetime is $\tau(bb\bar u\bar d) \sim 367$ fs.
The $T(bb\bar u\bar d)$ tetraquark can decay through one of two channels:
\nl
(a) The ``standard process" $bb \bar u \bar d \to c b \bar u \bar d +
W^{*-}$.
Typical reactions include \vrule width 0pt height 2.0ex depth 1.5ex
    $T(bb\bar u\bar d) \to D^0 \bar B^0 \pi^-$,\, $D^+ B^- \pi^-$
\ and
\ $T(bb\bar u\bar d) \to \jpsi K^- \bar B^0$,\, $\jpsi \bar K^0 B^-$.
In addition, there is a rare process where {\em both} $b$ quarks decay into
$c\bar c s$,
\vrule width 0pt height 2.5ex
   $T(bb\bar u\bar d) \to \jpsi \jpsi K^- \bar K^0$.
The signature for events with two $\jpsi$'s coming from the same secondary
vertex might be sufficiently striking to make it worthwhile to look for
such events against a large background.
\nl
(b) The $W$-exchange process $b \bar d \to c \bar u$, involving either
one of the two $b$ quarks.  The latter process can
involve a two-body final state, e.g., \,$T(bb\bar u\bar d) \to D^0 B^-$.

In contrast with \,$T(bb\bar u\bar d)$, 
the mass of \,$T(cc\bar u\bar d)$\, 
with $J^P{=}1^+$ is predicted to be $3882\pm12$ MeV, 7 MeV {\em above}
the $D^0 D^{*+}$ threshold and 148 MeV above $D^0 D^+ \gamma$ threshold.
$T(bc\bar u\bar d)$ with $J^P{=}0^+$ is predicted at $7134\pm13$ MeV, 11
MeV below the $\bar B^0 D^0$ threshold.  The theoretical 
precision is not sufficient to determine whether $bc\bar u\bar d$\, 
is actually above or below the threshold.
It could manifest itself as a narrow resonance just at threshold.

At this point it is interesting to point out an interesting pattern:
the known candidates for hadronic molecules are hidden-flavor
quarkonium-like states $Q\bar Q q \bar q$, $Q=b,c$, $q=u,d$,
while the stable tetraquark belongs to the open-flavor
$QQ\bar q\bar q$ category. There is a good reason for this pattern.

$T(bb \bar u \bar d)$ is below two-meson threshold because the two heavy quarks
are very close to each other ${\sim}0.2$ Fermi. They form a color
antitriplet and attract each other very strongly. Consider a typical
Coulomb + linear Cornell-like potential $V(r)={-}\alpha_s/r + \sigma r$. 
At ${\sim}0.2$ Fermi the heavy quarks probe
the Coulomb, singular part of the potential, so the binding energy is
very large, ${\sim}280$ MeV. But the tightly-bound $(bb)$ sub-system is 
a color antitriplet, so it cannot disconnect from the two light antiquarks. 
Hence the tetraquark is bound vs. two heavy-light $b\bar q$ mesons 
which lack the strong attraction between the two heavy quarks.

The situation is completely different in bottomonium-like system
$(b \bar b q \bar q)$: the lowest energy configuration of the
$(b \bar b)$ subsystem is a color singlet. So
when $b$ and $\bar b$ get close, they decouple from the light quarks and form
an ordinary bottomonium. In other words,
in a $(b \bar b q \bar q)$ system there is no possibility of utilizing 
the very strong attraction between $b$ and $\bar b$ without at the same time
forcing the system to decay into quarkonium and pion(s).

This is why exotic bottomonium-like states have a completely different
structure -- they are hadronic molecules of two heavy-light mesons bound
by exchange of light hadrons.  Such molecules have a mass which is much higher
than their decay products:  For example,
$Z_b(10610)$ is ${\sim}1$ GeV above $\Upsilon(1S) \pi$ threshold. 
Nonetheless, they have a strikingly narrow width despite such a large
phase space, e.g.,  $\Gamma(Z_b(10610))\sim 20$ MeV \cite{Garmash:2014dhx}.
The reason is that in order to decay into quarkonium and a pion
the two heavy quarks must get very close to each other.
In a large deuteron-like molecular state the probability for
such a close encounter is quite small, analogous to the small
probability for an electron to be inside the proton in the ground
state of a hydrogen atom.

Analogous comments apply to $cc\bar q\bar q$ states vs. $c\bar c q \bar q$,
states, with an important difference that $m_c/m_b\sim 1/3$, so the 
substantial binding energy of the $(cc)$ subsystem is nevertheless
significantly smaller that in $(bb)$ subsystem and therefore 
$c c \bar q \bar q$ is likely unbound with respect to two $(c\bar q)$
mesons.

\section{SUMMARY AND OUTLOOK}

\label{sec:summary}

The quark model has been highly successful in describing the spectroscopy
of mesons and baryons as quark-antiquark ($q \bar q$) and three-quark ($qqq$)
systems, respectively.  With the $u,d,s$ quarks assigned the fractional
charges $2/3,-1/3,-1/3$, these are the simplest states with integral charges.
However, the model also implied the existence of more complicated states with
integral charges, such as $qq \bar q \bar q$ mesons and $qqqq \bar q$ baryons
\cite{GellMann:1964nj,*Zweig:1981pd}.  Regarding quarks as fundamental triplets
of a color SU(3) symmetry, the color-singlet states have integral charges.
Why weren't these ``exotic'' states seen?

One signal of an exotic hadron is its ``flavor'' quantum numbers, calculated
from the charge $(2/3,-1/3,-1/3)$ and strangeness $(0,0,-1)$ of the $u,d,s$
quarks.  Thus, a meson with the quantum numbers of $uu \bar s \bar s$, decaying
to $K^+ K^+$ or $p \bar \Xi^+$, would be manifestly exotic, as its charge and
strangeness could not be exhibited by any $q \bar q$ state.  Similarly, a
baryon with the quantum numbers of $uudd \bar s$, decaying to $K^+ n$ or $K^0
p$, would be manifestly exotic.

Various models implied that exotic states of $u,d,s$ quarks existed, but were
not identifiable either because they did not possess exotic flavor quantum
numbers or because they were too broad to be distinguishable from two-hadron
continuum states.  A model in which quarks were confined by a
quantum-chromodynamics ``bag'' \cite{Jaffe:1976ig,*Jaffe:1976ih} predicted a
$q q \bar q \bar q$ meson as light as a few hundred MeV but with large decay
width to $\pi \pi$.  Narrower exotic mesons now known as $f_0$ and $a_0$ were
expected with masses about a GeV, as seen, but their flavor quantum numbers
are indistinguishable from those of $q \bar q'$ states.  In the bag model they
possess an additional $s \bar s$ pair in their wave functions, and thus are
known as ``crypto-exotic.''  The application of
quark-hadron duality to baryon-antibaryon scattering implied that $t$-channel
exchanges of (non-exotic) $q \bar q$ states were dual to (exotic) $qq \bar q
\bar q$ states in the $s$ channel \cite{Rosner:1968si}.  Thus, one expected
exotic states in baryon-antibaryon channels, such as $\Delta^{++}
\bar n$ and $\bar \Lambda p \pi^+$.  Their absence may be ascribed to
their large decay widths.

The situation has changed with the advent of the heavy charm ($c$) and bottom
($b$) quarks.  A multitude of exotic hadrons with two or more heavy quarks
have been seen, starting with the $X(3872)$ \cite{Choi:2003ue}, where the
number in parentheses refers to the mass in MeV.  Several mechanisms appear to
be at work in these observations.  In the case of the $X(3872)$,
one-pion-exchange between a charmed meson $D$ and an anti-charmed meson
$\bar D^*$, binding them into a bound or virtual S-wave state, plays
a crucial role.  The spin $J$, parity $P$, and charge-conjugation
eigenvalue $C$ of the state are then expected to be $J^{PC} = 1^{++}$, as
observed.  The isospin splitting between neutral and charged charmed mesons
ensures that the meson-antimeson component of the $X$ wave function is mainly
$D^0 \bar D^{*0}$, so it is a mixture of isospins zero and one with quark
content $c \bar c u \bar u$.  Its decay to $J/\psi \gamma$ and $\psi(2S)
\gamma$ implies that it has some $c \bar c$ in its wave function.  It would
then be a mixture of a molecular state and the first radial excitation of the
$\chi_{c1}(1P)$ state.  [The notation is that of Ref.\ \cite{Olive:2016xmw}].

The name ``tetraquark'' conventionally refers to a state in which all quarks
and antiquarks participate democratically in binding.  For the $X(3872)$ to
be identified as a tetraquark, the grouping into a charmed meson-antimeson
pair has to be ignored, and there has to be a charged partner with the same
$J^P$ nearby in mass.  The $Z_c(3900)$ would have been a possible candidate
except that it has $C = -$ instead of $C = +$ \cite{Eidelman:2016}.  In the
absence of full isospin multiplets, one cannot yet identify many exotic hadrons
as tetraquarks or pentaquarks.

The bottom counterpart of the $X(3872)$ has yet to be identified.  It may
participate in some mixing with a state thought to be the $\chi_{b1}(3P)$
\cite{Karliner:2014lta}.  Because its constituent $B$ and $\bar B^*$ mesons
enjoy little isospin splitting, its isospin is expected to be mainly zero, so
it should decay to $\Upsilon(1S,2S,3S)\kern0.05em\omega$, unlike $X(3872)$
which decays both to $\jpsi\omega$ and $\jpsi\rho$ \cite{MKtoCMS,*Guo:2014sca}.

Strong candidates for molecular states also exist in the bottom sector.  The
masses of $Z_b(10610)$ and $Z_b(10650)$ are very close to the $B \bar B^*$ and
$B^* \bar B^*$ thresholds, respectively.  They are seen not only in the
$\Upsilon(1S,2S,3S) \pi$ channels (only charged pion for $Z_b(10650)$),
but also in the $h_b(1P,2P) \pi^\pm$ channels, implying a violation of
heavy-quark symmetry \cite{Bondar:2011ev}.  This is to be expected if the
wave functions of the states are mainly meson-antimeson.  The important role
of pion exchange in creating these states is supported by the absence of
states near $B \bar B$ threshold.

A number of exotic meson candidates with $c \bar c$ accompanied by light
quarks have been seen in the 4--5 GeV mass range.  Many of these cannot
be associated with specific thresholds, and their tetraquark interpretation
often awaits discovery of their isospin counterparts.  In channels where pion
exchange is not possible, the role of $\eta$ exchange remains to be tested
\cite{Karliner:2016ith}.

A prominent feature of charmed-pair production by $e^+ e^-$ collisions is its
rapid drop just above a center-of-mass energy of 4.2 GeV, recalling similar
behavior in the $\pi \pi$ $I=J=0$ channel just around $K \bar K$ threshold.
In the charm case the behavior is likely to be correlated with the threshold
for $D(2420)\bar D$ production, which is the lowest-lying charmed pair which
can be produced in an S wave.  It illustrates the importance of S-wave
thresholds, which appear in a wide variety of cases in particle physics and
elsewhere \cite{Feshbach:1958nx,*Feshbach:1962ut,Rosner:2006vc}.

After a couple of false starts (by others) in searches for $qqqq \bar q$
states, the LHCb experiment has observed two in the $J/\psi p$ channel,
produced in the decay $\Lambda_b \to J/\psi K^- p$:  a narrow one around 4450
MeV and a much broader one around 4380 MeV, 
with opposite parities and preference for 3/2 and 5/2 spins, in either order.
A $\Sigma_c \bar D^*$ molecule with properties consistent with the
$P_c(4450)$ state has been suggested \cite{Karliner:2015ina}, 
but a molecular interpretation of the lower,
broader, state is elusive.  A genuine pentaquark interpretation would imply the
states are accompanied by numerous isospin partners, not yet observed.

A general feature of exotics with two or more heavy quarks is the reduction in
kinetic energy afforded by their large masses.  This, together with their
shorter Compton wavelength leading to deeper binding, implies that states
incorporating those heavy quarks may be deeply enough bound to overcome the
tendency to ``fall apart'' exhibited by exotic states composed only of
$u,d,s$ quarks.  An extreme example of this is the prediction of a bound
$b b \bar u \bar d$ state \cite{Karliner:2017qjm,Eichten:2017ffp}, supported
by methods used in the successful prediction of the mass of a baryon containing
two charm quarks \cite{Karliner:2014gca}.

Looking back at the experimental developments in hadron spectroscopy in
the new millennium:  heavy quarks have done it again! After converting us
into firm believers of the quark model in the seventies, heavy quark systems
have more recently taught us a new lesson: not all hadronic states are the
minimal quark combinations. In addition to $q\bar{q}$ mesons, four-quark
$qq\bar{q}\bar{q}$ configurations become important, especially near and
above the $q\bar{q}$ plus $q\bar{q}$ meson thresholds. Similarly, not all
baryons are $qqq$ states; $qqq\kern0.05em Q\bar{Q}$ configurations also play a
role.  Theoretical disputes rage on, if the observed multiquark configurations
are tightly bound tetra- and penta-quarks, or loosely bound meson-meson
and baryon-meson molecules. In our opinion, the case for the latter is
stronger.  It is also beyond any dispute that baryon-baryon molecules
exist and have been known for a long time as nuclei. This does not imply
that every multiquark system must be loosely bound. In fact, the models
which work well for doubly-charmed baryons also predict a stable $b b
\bar{u}\bar{d}$ tightly bound tetraquark.

What does the future hold for exotic multiquark mesons and baryons?  As
mentioned, the photoproduction of $J/\psi p$ resonances is possible at JLAB.
Production of charmonium-like states is envisioned at PANDA. We are likely to
be surprised by more charmonium-like exotics from Belle II and LHCb. After
its upgrades, the LHCb may have a shot at the $b b \bar{u}\bar{d}$
tetraquark. We are looking forward to these developments!

\section*{DISCLOSURE STATEMENT}
The authors are not aware of any affiliations, memberships, funding, or
financial holdings that might be perceived as affecting the objectivity of
this review. 

\section*{ACKNOWLEDGMENTS}
This work was supported by the National Science Foundation (USA) Award Number 1507572.

\ifx\mcitethebibliography\mciteundefinedmacro
\PackageError{LHCb.bst}{mciteplus.sty has not been loaded}
{This bibstyle requires the use of the mciteplus package.}\fi
\providecommand{\href}[2]{#2}

\end{document}